\documentclass[preprint,12pt]{elsarticle}

\usepackage{amssymb}

\usepackage{lineno}
\usepackage[hidelinks]{hyperref}
\hypersetup{colorlinks,bookmarksopen,linkcolor=blue,citecolor=blue,urlcolor=blue}
\usepackage{xcolor}
\usepackage{printlen}
\usepackage{float}
\uselengthunit{mm}

\journal{Wiley}

\usepackage[english]{babel}
\usepackage[utf8x]{inputenc}
\usepackage[T1]{fontenc}
\usepackage{bm}

\usepackage{pgfplotstable}
\usepackage{pgfplots}
\usepackage{tikz}
\usepackage{stanli}
\usepackage{amsmath}
\usepackage{amsfonts}
\usepackage{url}
\pgfplotsset{compat = 1.15}
\usepackage{graphicx}
\usepackage{color}
\usepackage{geometry}
\usepackage{setspace}
\usepackage{siunitx}
\usepackage{multirow}
\usepackage{graphicx}
\usepackage{changepage}
\usepackage{chngcntr}

\usetikzlibrary{external}
\usetikzlibrary{angles,quotes}

\usepackage{rotating}
\newcommand{\verttext}[1]{\begin{turn}{90}#1\end{turn}}
\usepackage{tabularray}
\usepackage{placeins}

\usepackage{pgfplots}
\usepackage{pgfplotstable}
\pgfplotsset{compat=newest}

\pgfplotsset{
  scaled tick label/.code={
    \pgfmathparse{10^(#1)}%
    \edef\pgfplotsretval{\ensuremath{\times 10^{\pgfmathprintnumber{#1}}}}%
  }
}

\usepgfplotslibrary{external}
\begin{document}
\emergencystretch 3em

\begin{frontmatter}



\title{Non-uniform pneumatic actuation switches macroscopic properties of elastomeric honeycombs}


\author[ctumech]{Ondřej Faltus\corref{cor1}}
\ead{ondrej.faltus@cvut.cz}
\cortext[cor1]{Corresponding author}
\author[ctumech]{Martin Doškář}
\ead{martin.doskar@cvut.cz}
\author[ctuec]{Jan Havelka}
\ead{jan.havelka@fsv.cvut.cz}
\author[ctuec]{Pavel Rychnovský}
\ead{pavel.rychnovsky@fsv.cvut.cz}
\author[tuemom]{Ondřej Rokoš}
\ead{o.rokos@tue.nl}

\affiliation[ctumech]{
            organization={Department of Mechanics, Faculty of Civil Engineering, Czech Technical University in Prague},
            addressline={Thákurova 7}, 
            postcode={166 29},
            city={Prague 6},
            country={Czech Republic}}

\affiliation[ctuec]{
            organization={Experimental Centre, Faculty of Civil Engineering, Czech Technical University in Prague},
            addressline={Thákurova 7}, 
            postcode={166 29},
            city={Prague 6},
            country={Czech Republic}}

\affiliation[tuemom]{
            organization={Mechanics of Materials, Department of Mechanical Engineering, Eindhoven University of Technology},
            addressline={P.O. Box 513}, 
            postcode={5600MB},
            city={Eindhoven},
            country={The Netherlands}}

\begin{abstract}
Honeycomb microstructures with circular voids are well known to undergo pattern transformations under macroscopic strain loading. Depending on the biaxiality of the applied strain, they deform into three different patterns. Here we demonstrate that the same three patterns can be triggered also by pneumatic actuation of the voids, with resulting patterns depending on an applied pressurization scheme, i.e., the ratio of suction pressures  introduced in different voids within the microstructure. Our numerically obtained findings are experimentally validated on two finite size samples made of silicone rubber cast in a 3D printed mold. In numerical studies, we first showcase the evolution of homogenized stiffness and its anisotropy during the genesis of pneumatically-induced patterns. Our loading schemes result in macroscopic stiffness varying by a factor of two with loading direction, chosen by a reversible actuation independent of the load itself. Second, we study the effect of the pneumatic actuation on acoustic properties of the microstructure, finding the proposed method viable to trigger different acoustic bandgaps.
\end{abstract}



\end{frontmatter}


\section{Introduction}

Pattern-forming metamaterials are a subclass of mechanical metamaterials characterized by their microstructure exhibiting internal buckling, which results in appealing macroscopic behavior, such as a variable or negative Poisson's ratio, or control of wave propagation~\cite{Kochmann2017}. In response to external stimuli, such as compressive macroscopic strain, the buckling process transforms the microstructure into one or more possible geometrical patterns which exhibit vastly different macroscopic properties in comparison to the pre-bifurcation state. In their simplest form, the pattern-forming metamaterials are realized typically as two-dimensional, theoretically infinite sheets of a soft material, such as polymer rubber, perforated periodically by voids of various shapes~\cite{Kochmann2017}.

\begin{figure}[t]
    \centering
    \includegraphics{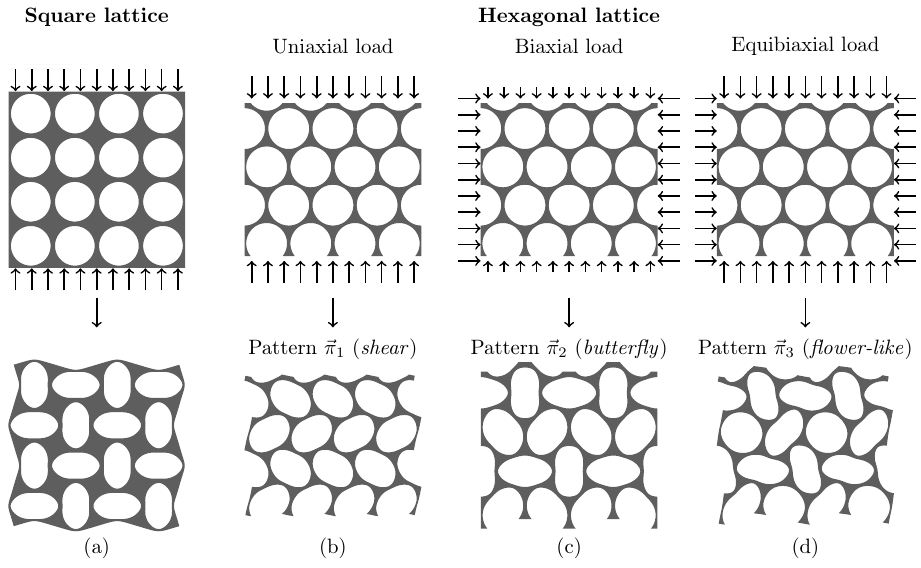}
    \caption{Instability patterns resulting from internal buckling of a microstructure comprising (a)~square stacking of circular voids, and (b-d)~hexagonal stacking of circular voids, under compressive macroscopic strain. Undeformed configurations and applied external loads are shown in the top row, and corresponding deformed configurations are plotted in the bottom row. 
    }
    \label{fig:knownpatterns}
\end{figure}

Among the basic examples is the square stacking of circular voids~\cite{Mullin2007patterning}. This arrangement features only one possible local buckling pattern under compressive strain: originally circular voids deform into ellipses with alternating horizontal and vertical direction of their major axes, see Figure \ref{fig:knownpatterns}a. The behavior is elastic and fully reversible, and leads to auxeticity, i.e., negative Poisson's ratio in the post-buckling state~\cite{Bertoldi2010auxeticityinpatterning}. A more complex response can be observed in the hexagonal stacking of circular voids, i.e., when the holes are arranged in a hexagonal lattice. This arrangement leads to three different buckling patterns under macroscopic compressive strain with the choice of pattern dictated by the biaxiality ratio of the applied loading, see Figures~\ref{fig:knownpatterns}b-d. Behavior of the hexagonal void arrangement was first studied experimentally by Papka and Kyriakides~\cite{papka1999honeycombcrushing, papka1999honeycombcrushinganalysis}. The bifurcating behavior and its stability was discussed extensively by Combescure~et~al.~\cite{combescure2016hexagonstabilityunderequibiaxial, combescure2020hexasquarestabilityofpatterns}. Much focus was also given to homogenizing this microstructural behavior~\cite{Ohno2002homogenizationonhoneycombs, Ameen2018sizeeffectonhoneycombs, Rokos2020honeycombs} and also to performing multi-scale buckling analysis~\cite{Johnson2017holeycolumn,Rokos2020newtonsolver}.

Mechanical behavior of these geometries can be further tuned. For instance, variations in void radii in a square lattice result in behavior programmable by lateral confinement~\cite{Florijn2014, Florijn2016geomrole}, or in shape memory effects~\cite{Yang2016shapememoryinsquarestack}. Introducing an interfacial layer into the microstructural ligaments can control the critical buckling strain as well as post-buckling material properties \cite{He2017compositesquarestack}. In addition, pattern-forming metamaterials are not limited to circular voids~\cite{Bertoldi2008Boycemat}. Tuning the void shape in a square lattice leads to high versatility in macroscopic response~\cite{Overvelde2012tuningholes,Overvelde2014poreshape}. Moreover, adding stiff coating on the void surfaces leads to different buckling patterns subject to thermal loads, resulting in several acoustic band gaps~\cite{Liu2022squarelatticecoatingbandgaps}.

Pattern-forming behavior can find applications in microstructure assembly through swelling processes~\cite{Zhang2008squarestackswelling} or through acrylic acid polymerization~\cite{Singamaneni2009frozenacidinvoids}. Francesconi et al.~\cite{Francesconi2019} demonstrated that layering honeycomb microstructures of varying geometry can lead to variable Poisson's ratio materials. Moreover, different lattice designs allow for reconfigurable auxetic and chiral materials~\cite{Shim2013auxeticchiralpatterning}, while different void shapes may be used as modules in modularly assembled microstructures~\cite{Doskar2023modularL}. Pattern transformation also enables tunable bandgaps and wave transmission properties~\cite{Bertoldi2008bandgapsinpatterning, Shan2014elasticwavesinhexagons, ning2021bandgapsinmetamatswithinclusions}. This is of special interest here since actively controlled acoustic metamaterials are a broadly developing research field~\cite{gorshkov2023bandgapsinmrelattice, jirasek2024integralmmbandgaps1, smejkal2024integralmmbandgaps2}.

While all the previous works relied on macroscopic strain, pneumatic actuation offers an alternative way to trigger the pattern formation. It has been studied numerically and experimentally primarily on square lattices, demonstrating programmability of the response through tuning of void size~\cite{Chen2018pneumaticpatterns} and usefulness of pneumatically induced patterning to trigger elastic band gaps~\cite{Liang2023pneusquarestackbandgap}. Pneumatic actuation can be exploited to control bifurcation of a novel patterning material with square voids~\cite{Hyatt2022pneumaticsquareholes}. A pneumatically actuated pattern-forming metamaterial has also been applied in soft robotics to control a simple gripper mechanism~\cite{Yang2015Bertoldigripper}. These studies limited themselves, however, to application of pneumatic actuation as a uniform suction load across all voids.

In our previous work, we demonstrated that non-uniform pneumatic loads can trigger internal patterning instabilities in square lattices of rectangular voids~\cite{Faltus2025squaremetamat}. In this contribution, we focus on pneumatic actuation of honeycomb microstructures. We present ways of triggering all three known patterns on the same microstructure by applying non-uniform air suction in various schemes, defined as distributions of suction pressure load magnitudes across different voids. To this end, we first employ finite element analysis on periodic representative volume elements (RVEs) to investigate which loading schemes trigger which patterns. Subsequently, we use numerical homogenization procedures to verify that the proposed way of triggering the patterns leads to macroscopic effective stiffness reduction and anisotropy of the macroscopic response in line with the patterns' known behavior under macroscopic strain loading. We continue with a numerical dispersion analysis to describe the acoustic behavior of the microstructure in response to the different schemes of pneumatic loading. Subsequently, we transfer the knowledge gained from this numerical analysis to an experimental setting, where we demonstrate the predicted behavior on finite samples made of silicone rubber.

The remainder of this article is structured as follows: Section \ref{sec:simulations} concerns the numerical simulations on RVEs of an infinite metamaterial and examination of the resulting macroscopic properties. In Section \ref{sec:experiments}, numerically obtained results are validated experimentally on finite structures, including direct numerical simulations of those structures to serve as a comparison between the numerical and experimental approach. Finally, Section \ref{sec:conclusions} offers concluding remarks.

\section{Pneumatic actuation schemes for honeycomb patterns}
\label{sec:simulations}

\subsection{Standard strain-triggered patterns}

\begin{figure}
    \centering
    \begin{tabular}{c c c}
         \includegraphics[width=.3\linewidth]{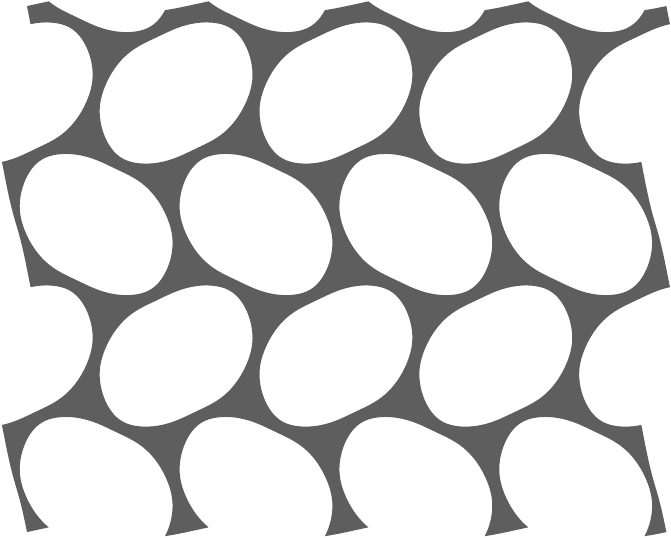}
         &
         \includegraphics[width=.3\linewidth]{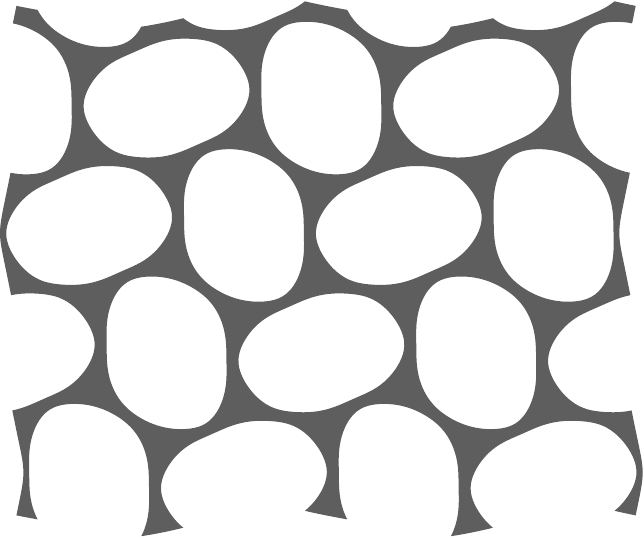}
         &
         \includegraphics[width=.3\linewidth]{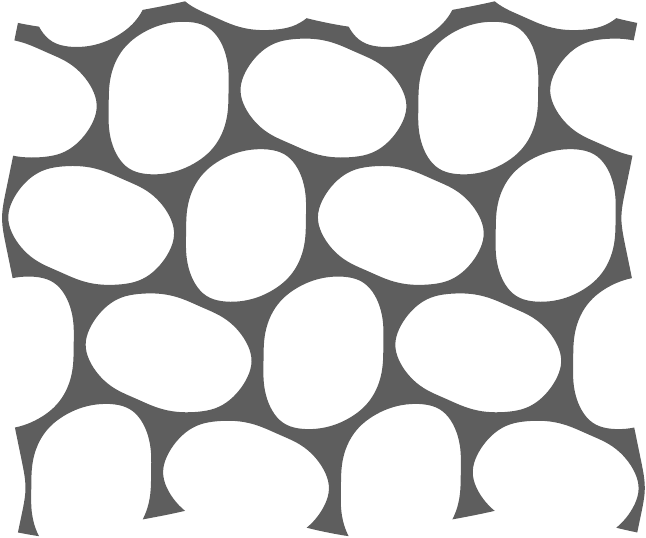}
         \\
         (a) {\footnotesize Mode $\vec{\varphi}_1$} & {\footnotesize (b) Mode $\vec{\varphi}_2$} & {\footnotesize (c) Mode $\vec{\varphi}_3$}
    \end{tabular}
    \caption{The different modes $\vec{\varphi}_i$, $i \in \{1,2,3\}$, of the pattern $\vec{\pi}_1$ on a hexagonal lattice of circular voids. These shear modes differ only in their orientation with regards to the symmetry axes od the reference geometry.}
    \label{fig:modes}
\end{figure}

Two-dimensional honeycomb microstructures with hexagonal stacking of circular voids collapse into one of three distinct patterns when subjected to compressive macroscopic strains~\cite{Ohno2002homogenizationonhoneycombs}. The formation of pattern is dictated by the biaxility ratio of the strain as indicated in Figure \ref{fig:knownpatterns}b-d. This ratio $\gamma$ can be defined as

\begin{equation}
\label{eq:strainratio}
    \gamma = \frac{|\bar{F}_{22}-1|}{|\bar{F}_{11}-1|} \in \langle 0, \infty )
\end{equation}

\noindent where $\bar{\bm{F}} = \bar{F}_{ij}\vec{e}_i\vec{e}_j$ is the macroscopic deformation gradient tensor, with $\vec{e}_i$ the base vectors of the coordinate space.

If the compressive strain is uniaxial along one of the three symmetry axes of the microstructure (for instance the vertical axis in Figure~\ref{fig:knownpatterns}b), pattern~$\vec{\pi}_1$, also known as the \textit{shear pattern}, arises. Under a general biaxial loading, where the strain along two symmetry axes (the inclined ones in Figure~\ref{fig:knownpatterns}c) is larger than strain along the remaining one, pattern~$\vec{\pi}_2$, the \textit{butterfly pattern}, appears. When the loading is exactly equibiaxial, the \textit{flower-like pattern}, $\vec{\pi}_3$, is triggered (Figure~\ref{fig:knownpatterns}d). This last pattern, however, is highly sensitive to perfect symmetry of both microstructure and loading, and thus even slight mesh asymmetries in simulation, or geometrical and loading imperfections in experiment, hamper its formation.

Pattern $\vec{\pi}_1$ has three modes, denoted as $\vec{\varphi}_i$, $i=1,2,3$, differing by the pattern's orientation, see Figure~\ref{fig:modes}. Interestingly, these modes form the linear basis of the pattern space, i.e., all of the patterns $\vec{\pi}_1$, $\vec{\pi}_2$, and $\vec{\pi}_3$ are their linear combinations~\cite{Rokos2020honeycombs}:

\begin{equation}
\label{eq:modelincombs}
    \vec{\pi}_1 = \vec{\varphi}_1 \quad\quad \vec{\pi}_2 = \vec{\varphi}_2 + \vec{\varphi}_3 \quad\quad \vec{\pi}_3 = \vec{\varphi}_1 + \vec{\varphi}_2 + \vec{\varphi}_3
\end{equation}

\subsection{Pressurization schemes for pattern triggering}
\label{sec:rveresults}

\begin{figure}[t]
    \centering
    \includegraphics{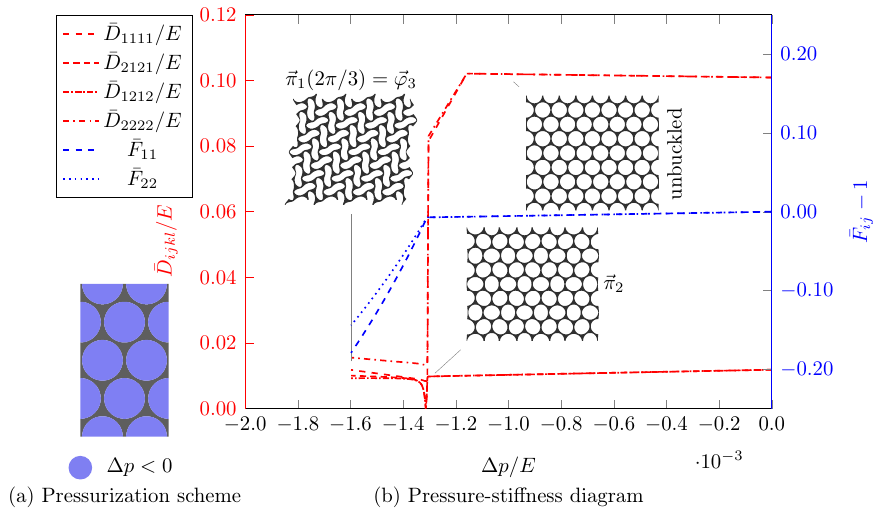}
    \caption{Pressurization scheme $s_0$ (constant) and the resulting evolution of the patterning process. (a)~Pressure setting. (b)~Evolution of selected macroscopic effective stiffness tensor components $\bar{D}_{ijkl}$ and macroscopic effective deformation gradient components $\bar{F}_{ij}$ with increasing pneumatic load $\Delta p < 0$. All stress and stiffness values are normalized by the bulk material Young's modulus $E$. Deformed states at various stages of loading are shown as insets in (b), depicting the shift in the microstructure between different internal patterns.}
    \label{fig:rveresult_constant}
\end{figure}

As discussed, under macroscopic strain loading the emergence of an internal pattern is dictated by the macroscopic strain ratio $\gamma$, which in turn dictates the load on the ligaments of the microstructure and thus the mode and order in which they buckle to form the buckling pattern. Here we replicate these microstructural conditions by designing non-uniform pressurization schemes of pneumatic loading. The investigation is conducted using plane strain finite element simulations of representative volume elements (RVEs) of infinite microstructures (see \hyperref[sec:numerical_methodology]{Methods} for more detail).

The simplest possible pressurization scheme is constant suction in all voids, which we denote as scheme $s_0$. This scheme resembles the equibiaxial loading condition known to produce the flower-like pattern $\vec{\pi}_3$. The scheme is pictured in Figure~\ref{fig:rveresult_constant}a, in which suction is represented as the negative value of $\Delta p$, the pressure difference with respect to the atmospheric pressure.

\begin{figure}
    \centering
    \includegraphics{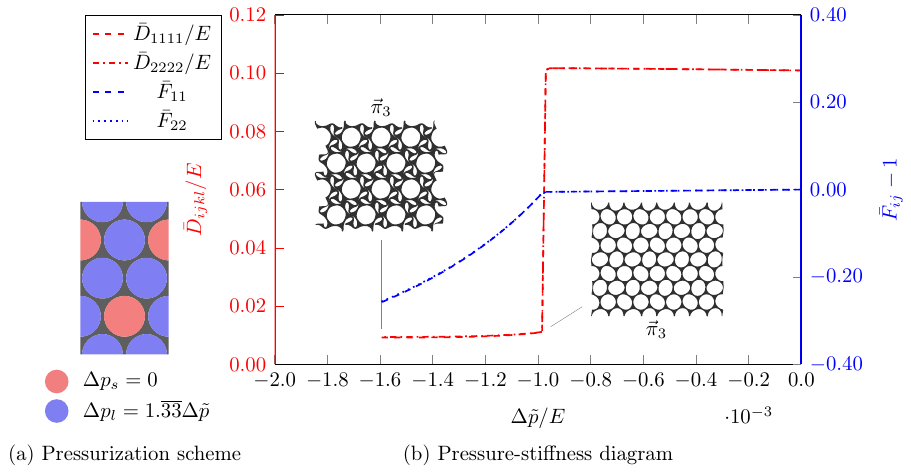}
    \caption{Pressurization scheme $s_1$ (constant with a pattern $\vec{\pi}_3$ modification) and the resulting evolution of the patterning process. (a)~Pressure setting. (b)~Evolution of selected macroscopic effective stiffness tensor components $\bar{D}_{ijkl}$ and macroscopic effective deformation gradient components $\bar{F}_{ij}$ with increasing pneumatic load $\Delta \tilde{p}$ . All stress and stiffness values are normalized by the bulk material Young's modulus $E$. Deformed states at and post-bifurcation are shown as insets in (b), depicting the patterning process.}
    \label{fig:rveresult_patternIII}
\end{figure}

Figure~\ref{fig:rveresult_constant}b shows the evolution of selected components of the effective macroscopic stiffness tensor $\bar{\mathbf{D}}$ with decreasing values of $\Delta p$. The bifurcation point is clearly visible as a drop in the depicted stiffness components at the critical load of $\Delta p_\mathrm{crit,1} = 1.304\cdot10^{-3}E$. Furthermore, another sharp corner can be identified from the graph of the shear stiffness components $\bar{D}_{2121}$ and $\bar{D}_{1212}$ almost immediately after at the critical load of $\Delta p_\mathrm{crit,2} = 1.313\cdot10^{-3}E$; this corner corresponds to a secondary pattern switch.

This twice bifurcating response is because the observed pattern is not the expected flower-like pattern $\vec{\pi}_3$, even though the strain ratio $\gamma$, as defined in Equation~(\ref{eq:strainratio}), stays close to $1$ (furthest away from $1$ reaching $\gamma = 1.0006$) in the entire pre-bifurcation regime. Instead, the microstructure first deforms towards the butterfly pattern $\vec{\pi}_2$ before eventually settling on the shear pattern $\vec{\pi}_1$. This evolution of deformed states is illustrated with insets at the corresponding points in the pressure-stiffness diagram in Figure~\ref{fig:rveresult_constant}b. Under the pneumatic loading, unlike the strain loading, the RVE is only simply supported and thus $\gamma$ is obtained from the equilibrium macroscopic deformation gradient $\bar{\bm{F}}$ while the macroscopic first Piola-Kirchhoff stress is $\bar{\bm{P}} = 0$. 

\begin{figure}
    \centering
    \includegraphics{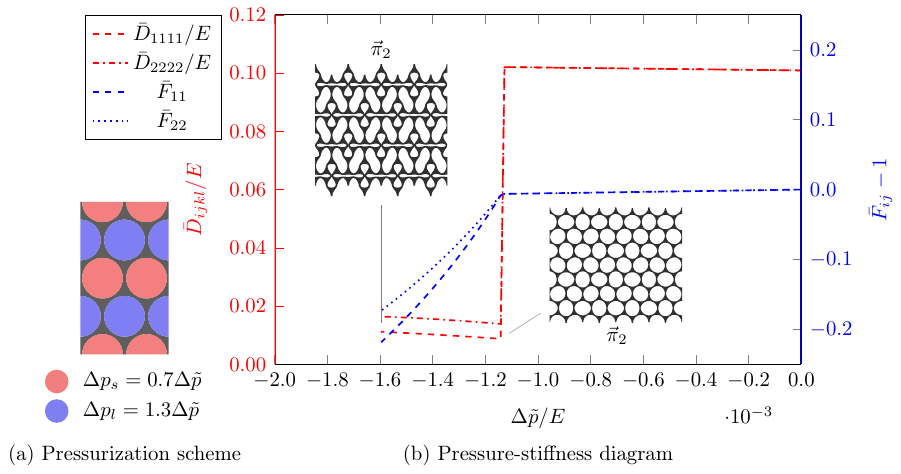}
    \caption{Pressurization scheme $s_2$ (row-wise) and the resulting evolution of the patterning process, leading to butterfly pattern $\vec{\pi}_2$. (a)~Pressure setting. (b)~Evolution of selected macroscopic effective stiffness tensor components $\bar{D}_{ijkl}$ and macroscopic effective deformation gradient components $\bar{F}_{ij}$ with increasing pneumatic load $\Delta \tilde{p}$. All stress and stiffness values are normalized by the bulk material Young's modulus $E$. Deformed states at and post-bifurcation are shown as insets in (b), depicting the patterning process.}
    \label{fig:rveresult_rows}
\end{figure}

We attribute this behavior to the sensitivity of pattern $\vec{\pi}_3$ to precise equibiaxial loading conditions and hence its susceptibility to imperfections including mesh non-symmetry~\cite{combescure2016hexagonstabilityunderequibiaxial}. Similar behavior has been previously reported for macroscopic strain loading cases with biaxiality ratios close but not exactly equal to $1$~\cite{Rokos2020honeycombs,Rokos2020newtonsolver}. During iterations of the modified Newton solver at the instability, any linear combination of all three patterns may occur and individual patterns are briefly observed in the unconverged deformed states before the algorithm finds a stable branch. Recall that $\gamma = 1$ exhibits a triple bifurcation point~\cite{Rokos2020newtonsolver} when loaded through externally applied macroscopic strain. Furthermore, a current-state eigenvalue analysis on the unstable branch (solved by a simple Newton solver without the Cholesky modification algorithm) confirms that three separate eigenvalues close in magnitude reach zero at the first bifurcation point. Their associated eigenmodes correspond to the three basis modes $\vec{\varphi}_i$, $i=1,2,3$ which the patterns are linear combinations of, recall Equation~(\ref{eq:modelincombs}). Moreover, there are multiple possible configurations corresponding to the $\vec{\pi}_3$ pattern on the RVE, differing by the position of the central void of the flower (i.e., a translation, rotation, and mirroring of the pattern).

\begin{figure}[t]
    \centering
    \includegraphics{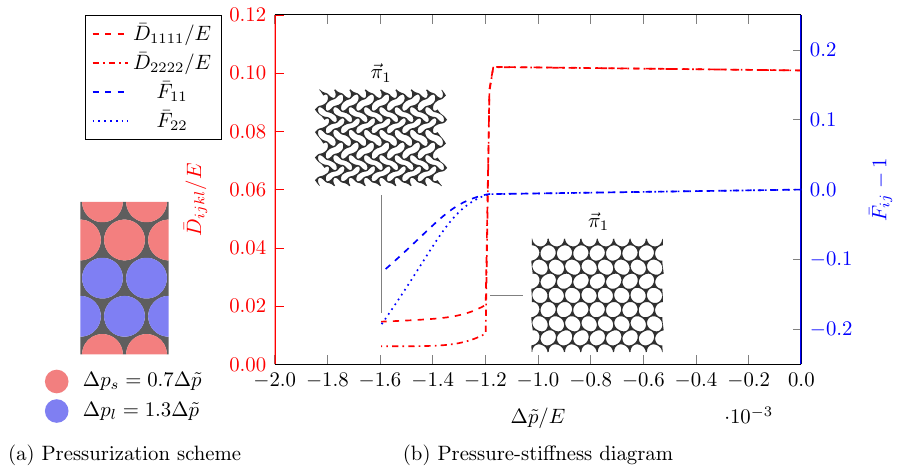}
    \caption{Pressurization scheme $s_3$ (double-row-wise) and the resulting evolution of the patterning process, leading to the shear pattern $\vec{\pi}_2$. (a)~Pressure setting. (b)~Evolution of selected macroscopic effective stiffness tensor components $\bar{D}_{ijkl}$ and macroscopic effective deformation gradient components $\bar{F}_{ij}$ with increasing pneumatic load $\Delta \tilde{p}$. All stress and stiffness values are normalized by the bulk material Young's modulus $E$. Deformed states at and post-bifurcation are shown as insets in (b), depicting the patterning process.}
    \label{fig:rveresult_doublerows}
\end{figure}

For that reason we propose an alternative pressurization scheme $s_1$, shown in Figure~\ref{fig:rveresult_patternIII}a, to achieve pattern $\vec{\pi}_3$. While the majority of voids remains depressurized uniformly (blue in Figure~\ref{fig:rveresult_patternIII}a), the two voids pre-selected as the centers of the "flower" in the pattern are not pressurized at all (red in Figure~\ref{fig:rveresult_patternIII}a). Therefore, two pressure differences are introduced: $\Delta p_s = 0$ and $\Delta p_l = 1.\overline{33}\Delta \tilde{p}$, where $\Delta \tilde{p}$ is the average pressure difference across all voids. This pressurization scheme leads to a very similar strain ratio of $\gamma=1.0006$ in the immediate pre-bifurcation state. Figure~\ref{fig:rveresult_patternIII}b shows again the evolution of macroscopic stiffness components in relation to $\Delta \tilde{p}$. Only one critical load is visible now, at the point where the microstructure forms the expected flower-like pattern $\vec{\pi}_3$ pictured in the insets in Figure~\ref{fig:rveresult_patternIII}b.

To trigger the butterfly pattern $\vec{\pi}_2$ by macroscopic strain, loading conditions need to be devised such that biaxial, yet not equibiaxial macroscopic strain is introduced. We propose to mimic this on the microstructural level with the pneumatic actuation scheme $s_2$, a row-wise alternation of smaller and larger pneumatic load magnitudes in the voids, see Figure~\ref{fig:rveresult_rows}a. Yet again the two levels of load, $\Delta p_s = 0.7\Delta \tilde{p}$ and $\Delta p_l = 1.3\Delta \tilde{p}$, are expressed as multiples of their average value $\Delta \tilde{p}$. This scheme succeeds to introduce pattern $\vec{\pi}_2$, as can be seen in the insets of Figure~\ref{fig:rveresult_rows}b. The associated macroscopic effective stiffness evolution is pictured in Figure~\ref{fig:rveresult_rows}b, with a single bifurcation point clearly visible. The strain ratio immediately pre-bifurcation is, nevertheless, still very close to $1$ in this case, $\gamma = 1.0032$.

Albeit the shear pattern $\vec{\pi}_1$ has already appeared as the secondary instability in the constant pressurization scheme $s_0$, it remains to be triggered directly and robustly. We find this to be possible with the pressurization scheme $s_3$, displayed in Figure~\ref{fig:rveresult_doublerows}a. The strain ratio induced by this actuation scheme is $\gamma = 1.0155$, slightly larger than for the previous schemes yet still close to $1$. The observed post-bifurcation pattern is, however, consistent with patterns induced by macroscopic strain loads of $\gamma > 1$ (compare insets of Figure~\ref{fig:rveresult_doublerows}b and Figure~\ref{fig:knownpatterns}b).

This concludes the definition of pressurization schemes that can trigger the three known patterns $\vec{\pi}_1$, $\vec{\pi}_2$, and $\vec{\pi}_3$. It remains to be seen whether those are, despite visual resemblance, actually mechanically the same as their macroscopic-strain-triggered counterparts in Figure~\ref{fig:knownpatterns}b-d, however. We remark that especially in the pre-bifurcation state, the behavior of the microstructure can exhibit differences compared to the strain-loaded case, as the presence of the pneumatic actuation in the voids introduces additional stress loading. The deformed state of the microstructure under pneumatic loading thus also exhibits minor correlation with the long wavelength modes $\vec{\varphi}_i$, $i=1,2,3$ (see Figure~\ref{fig:modes}) even prior to the bifurcation point.

\subsection{Effect of patterning on macroscopic stiffness}
\label{sec:rveproperties}

\begin{figure}[t]
    \centering
    \includegraphics{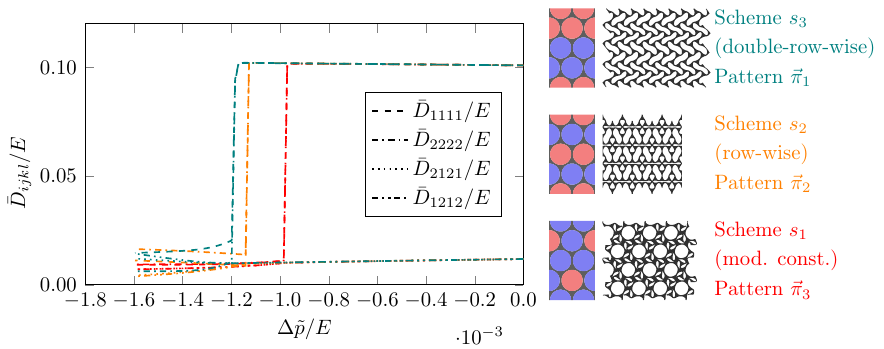}
    \caption{Evolution of the normal components $\bar{D}_{1111}$ and $\bar{D}_{2222}$ and the shear components $\bar{D}_{2121}$ and $\bar{D}_{1212}$ of the effective macroscopic stiffness tensor in response to pneumatic actuation by average suction pressure difference $\Delta \tilde{p}$ introduced via different pressurization schemes and thus triggering different internal patterns. All stress and stiffness values are normalized by the bulk material Young's modulus $E$.}
    \label{fig:stiffnesscomparison}
\end{figure}

The possibility to trigger all three known patterns of a hexagonal lattice by pneumatic actuation allows for pre-selecting how a single microstructure responds to macroscopic stimuli. Moreover, this selection is facilitated by pneumatic pre-actuation and thus entirely independent of the macroscopic strain loading. Figure~\ref{fig:stiffnesscomparison} compares the evolution of the normal components of the macroscopic stiffness tensor $\bar{\mathbf{D}}$ pre- and post-bifurcation for each of the patterns triggered. Admittedly, there is not a significant difference in the stiffness values among the different pressurization schemes/triggered patterns, with the stiffness values immediately post-bifurcation approximately at one tenth of its reference value in all cases. There is, however, a significant difference in other aspects of the post-bifurcation behavior.

For instance, patterns $\vec{\pi}_1$ and $\vec{\pi}_3$ lead to softening with increasing pneumatic load, while pattern $\vec{\pi}_2$ causes hardening in the vertical stiffness component ($\bar{D}_{2222}$ is increasing with decreasing applied pressure difference $\Delta \tilde{p}$), even without self contact coming into play.

\begin{figure}[t]
    \centering
    \includegraphics{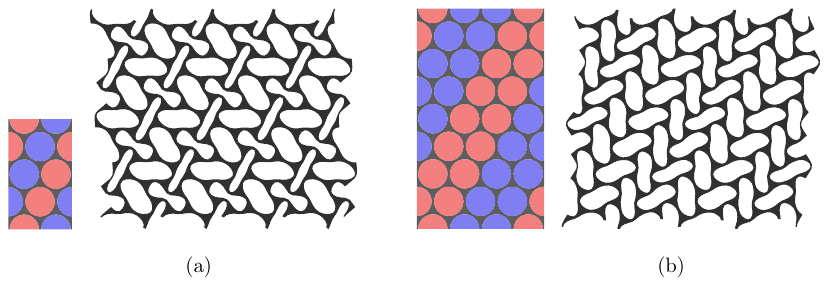}
    \caption{Patterns triggered by rotated pressurization schemes. (a) The row-wise pressurization scheme $s_2$ rotated by $\pi/3$, and the resulting post-bifurcation butterfly pattern. (b) The double-row-wise pressurization scheme $s_3$ rotated by $\pi/3$ and the resulting post-bifurcation shear pattern. This corresponds to the $\vec{\varphi}_2$ mode of pattern $\vec{\pi}_1$. A larger size RVE has to be used in this case to accommodate the periodicity requirements of the pressurization scheme.}
    \label{fig:rotateddeformations}
\end{figure}

Furthermore, the post-bifurcation behavior is isotropic for pattern $\vec{\pi}_3$, while anisotropic for the other patterns. In this context it is useful to note that due to the inherent symmetries of the hexagonal lattice, the pressurization schemes lacking these symmetries, i.e., the $s_2$ row-wise and the $s_3$ double-row-wise schemes, see Figures~\ref{fig:rveresult_rows}a and \ref{fig:rveresult_doublerows}a, can also be applied rotated by $\pm {\pi}/3$ with respect to the orientation presented above. The scheme orientation influences the orientation of the post-bifurcation pattern and thus the direction of the anisotropy of the resulting behavior. For pattern $\vec{\pi}_1$, this essentially means that all its modes $\vec{\varphi}_i$, $i=1,2,3$, can be triggered. Figure~\ref{fig:rotateddeformations} showcases the rotated schemes for the $+\pi/3$ case. Note that for the double-row-wise pressurization scheme $s_3$, these rotated schemes necessitate the use of a larger RVE (see Figure~\ref{fig:rve_geometry}b in \hyperref[sec:numerical_methodology]{Methods}).

\begin{figure}[H]
    \centering
    \includegraphics{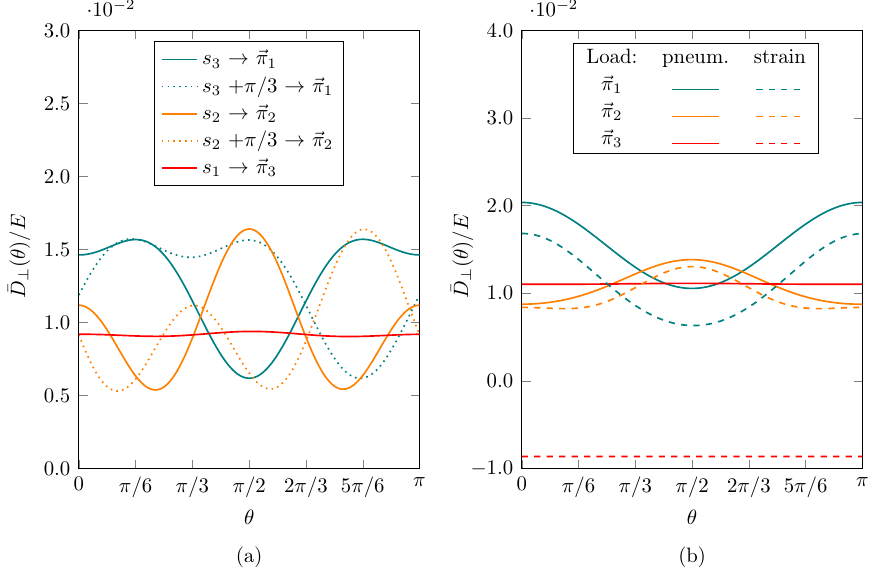}
    \caption{Macroscopic normal stiffness component $\bar{D}_\perp$ (relative to bulk material Young's modulus $E$) in the post-bifurcation state  projected into different directions rotated by angle $\theta$ (oriented counterclockwise, with $\theta=0$ at the $\vec{e}_1$ direction). (a) Computed at $\Delta \tilde{p} = 1.6\cdot10^{-3}E$ for all pneumatic loading schemes. The response of pattern $\vec{\pi}_3$, obtained from the modified constant pressurization scheme $s_1$, is almost isotropic. Dotted lines represent results for schemes rotated by $+\pi/3$, see Figure~\ref{fig:rotateddeformations}. (b) Computed immediately post-bifurcation for both pneumatically triggered patterns (solid lines) and patterns triggered by macroscopic strain loading (dashed lines).}
    \label{fig:isotropy}
\end{figure}

The (an)isotropy of the post-bifurcation behavior can be assessed through a rotated projection $\bar{D}_\perp$ of the normal component of macroscopic effective stiffness in the post-bifurcation state defined as

\begin{equation}
    \bar{D}_\perp(\theta) = \bm{A}(\theta) : \bar{\mathbf{D}} : \bm{A}(\theta)
\end{equation}

\noindent where $\bm{A}(\theta) = (\cos{\theta}\;\vec{e}_1 + \sin{\theta}\;\vec{e}_2) \otimes (\cos{\theta}\;\vec{e}_1 + \sin{\theta}\;\vec{e}_2)$, as illustrated in Figure~\ref{fig:isotropy}. The angle $\theta$ corresponds to the $\vec{e}_1$ direction at $\theta = 0$ and spans the interval $\langle 0,\pi \rangle$. For the row-wise and double-row-wise scheme, the results of the rotated schemes from Figure~\ref{fig:rotateddeformations} are also shown in Figure~\ref{fig:isotropy}a, predictably shifted by $\pi/3$ compared to their unrotated counterparts. 

In Figure~\ref{fig:isotropy}b, we additionally show a comparison between this post-buckling anisotropy in the presently studied pneumatically loaded case and in a case where the patterns are triggered by equivalent macroscopic strain loading (see Figure~\ref{fig:knownpatterns}b-d for illustration). Despite different magnitudes of the post-buckling stiffness, the anisotropy trends do qualitatively correspond to each other for each apparent pattern. Of interest is the situation for pattern $\vec{\pi}_3$, which is unstable (with negative macroscopic stiffness) under strain loading, as previously observed~\cite{combescure2016hexagonstabilityunderequibiaxial}. In the pneumatic case, however, the presence of the pneumatic load enforcing the pattern stabilizes it.


\begin{figure}[t]
    \centering
    \includegraphics{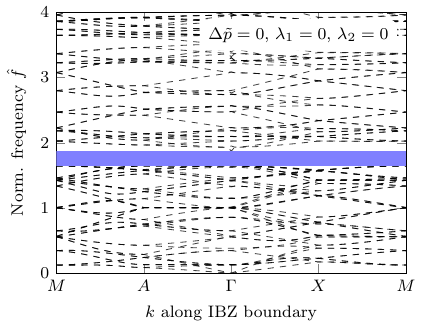}
    \caption{Dispersion diagram for wave vectors along the boundary of the Irreducible Brillouin Zone of the microstructure in the reference configuration.}
    \label{fig:referencedispersion}
\end{figure}

\subsection{Effect of patterning on acoustic properties}
\label{sec:bloch}

In addition to the study of post-buckled macroscopic stiffness, here we investigated the effect of pneumatically actuated patterns on the effective acoustic wave transmission properties of the microstructure. For the three pressurization schemes $s_1$, $s_2$, and $s_3$, we performed dispersion analysis~\cite{cool2024dispersionfordummies} throughout the whole pneumatic loading cycle. Additionally, for comparison, we did the same also for macroscopic strain loading of the same unit cell. A detailed description of the methodology is provided in \hyperref[app:bloch]{Methods}.

In reference configuration, the acoustic properties of the RVE are, naturally, the same in all cases. The dispersion diagram is pictured in Figure~\ref{fig:referencedispersion}. A singular bandgap at the $\hat{f}_\mathrm{gap} = \langle 1.651, 1.872 \rangle$ frequency range is featured (see Equation~(\ref{eq:freqnorm}) in \hyperref[app:bloch]{Methods} for the definition of $\hat{f}$).

\begin{figure}[p]
    \centering
    \includegraphics[width=\textwidth]{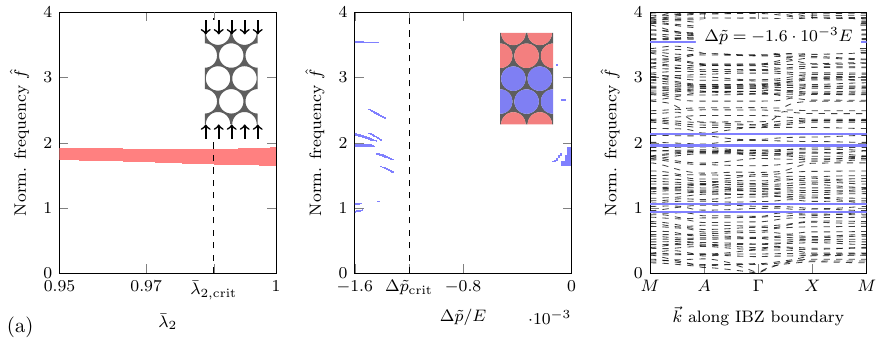}
    \includegraphics[width=\textwidth]{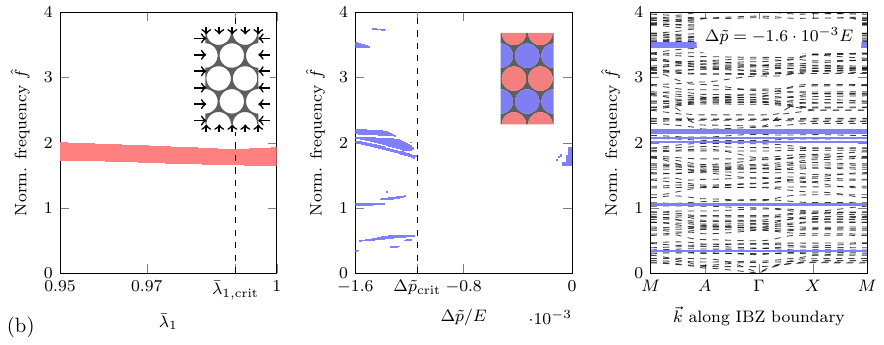}
    \includegraphics[width=\textwidth]{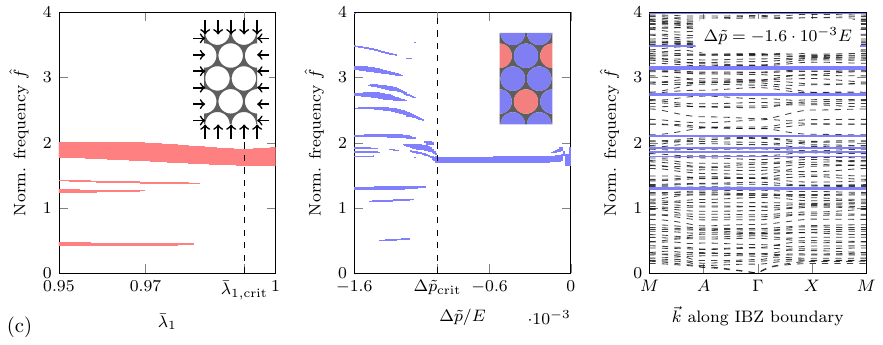}
    \caption{Dispersion analysis results for microstructure RVEs actuated both by macroscopic strain and pneumatically. Each row corresponds to one of the buckling patterns, loading schemes, and pressurization schemes. (a) $\vec{\pi_1}$: uniaxial load or scheme $s_3$, (b) $\vec{\pi_2}$: biaxial load or scheme $s_2$, and (c) $\vec{\pi_3}$: equibiaxial load or scheme $s_1$. Plots in the first column show bandgap evolutions during macroscopic compressive loading up to stretch $\bar{\lambda}_1=0.95$ or $\bar{\lambda}_2=0.95$. Second column shows bandgap evolutions in response to pneumatic loading  up to $\Delta \tilde{p} = - 1.6\cdot10^{-3}E$. Critical values of the load parameters are indicated by dashed black lines to relate the bandgap evolution to the patterning process. The last column shows dispersion diagrams for wave vectors along the boundary of the Irreducible Brillouin Zone (IBZ) for the pneumatic loading at the load level of $\Delta \tilde{p} = - 1.6\cdot10^{-3}E$.
    }
    \label{fig:blochresults}
\end{figure}

The evolution of this bandgap with loading is pictured in the first two columns of Figure~\ref{fig:blochresults}. The first column details what happens in response to macroscopic strain load, while the second column concerns pneumatic actuation up to $\Delta\tilde{p} = 1.6\cdot10^{-3}E$.

 The strain load results are not very remarkable. The reference bandgap is maintained throughout loading, while different smaller bandgaps open upon bifurcation in the flower-like buckling pattern $\vec{\pi}_3$. With pneumatic actuation of the RVE, however, the situation differs. Prior to any bifurcation, the reference bandgap disappears for the $s_2$ row-wise and $s_3$ double-row-wise pressurization schemes, while being maintained in the case of the $s_1$ modified constant pressurization scheme. We attribute this to the isotropy of the $\vec{\pi}_3$ pattern throughout the loading. Upon bifurcation into the microstructural patterns, numerous small bandgaps open for all pressurization schemes, see Figure~\ref{fig:blochresults} middle column for the evolution and right column for a snapshot of the dispersion diagram. The number and exact frequency ranges of these bandgaps differ among the pressurization schemes and bifurcation patterns, enabling possibly a fine-tuned choice of bandgap driven by selection of pressurization scheme. We nevertheless consider the aforementioned presence/disappearance of the reference bandgap in the pre-bifurcation stage to be the primary use case of pneumatic actuation for dynamic purposes.

\subsection{Overview of macroscopic effects}

Table~\ref{tab:rvesummary} summarizes induced patterns and differences in post-bifurcation behavior for all three pressurization schemes. The results show that within a theoretically infinite sheet, pneumatic actuation can serve as a method of switching between different modes of macroscopic behavior. In this way, it is possible to preselect the anisotropy of the metamaterial's response to macroscopic compression and its orientation even before macroscopic loading starts. Since the patterning behavior is a hyperelastic and reversible phenomenon, this switching can be performed repeatedly. In the same way, acoustic response of the material can also be controlled by the selection of the pneumatic actuation scheme and hence the occurence of various acoustic bandgaps.

\begin{sidewaystable}[p]
    \footnotesize
    \centering
    \begin{tblr}{|Q[c,h]Q[c,m]Q[c,m]|Q[c,h]Q[c,h]|Q[c,h]Q[c,h]Q[c,h]Q[c,h]|Q[c,h]Q[c,h]|}
        \hline
        & \SetCell[r=2,c=2]{l} Scheme & & \SetCell[c=2]{m} Patterning & & \SetCell[c=4]{m} Post-bifurcation macroscopic properties & & & & \SetCell[c=2]{m} Acoustic bandgaps
        \\ \cline{4-11}
        & & & Pattern & $\Delta \tilde{p}_\mathrm{crit}$ & $\bar{D}_{1111}^\mathrm{postbif}$ & $\bar{D}_{2222}^\mathrm{postbif}$ & $\theta_{\bar{D}_{\perp,\mathrm{min}}}$ & $\theta_{\bar{D}_{\perp,\mathrm{max}}}$ & Pre-bifur. & Post-bifur.
        \\ \hline \hline
         & \SetCell[r=2]{c,m} \verttext{Double-} \verttext{row-wise} & \SetCell[r=2]{c,m} \includegraphics[width=0.05\textwidth]{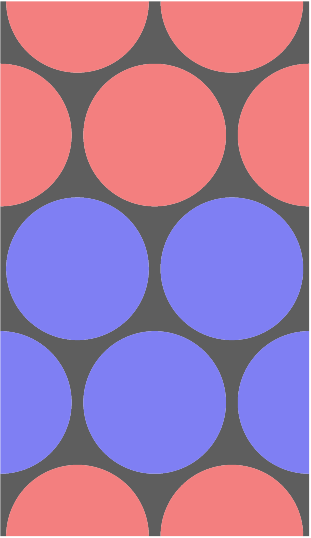} & & & & & &
        \\
        {$s_3$} & & & $\vec{\pi}_1$ & $-1.17\cdot10^{-3} E$ & {$0.020 E$} & {$0.011 E$} &  {$\pi/2$} & {$[\pi/6,5\pi/6$]} & None & Many small
        \\ \hline
         & \SetCell[r=2]{c,m} \verttext{Row-wise} & \SetCell[r=2]{c,m} \includegraphics[width=0.05\textwidth]{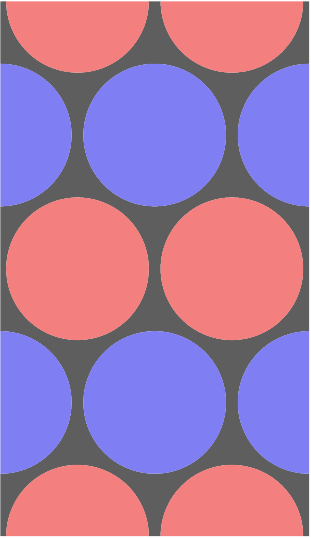} & & & & & &
        \\
        $s_2$ & & & {$\vec{\pi}_2$} & {$-1.13\cdot10^{-3} E$} & {$0.009 E$} & {$0.014 E$} &  {$[0.22\pi,0.78\pi]$} & {$\pi/2$} & None & Many small
        \\ \hline
        & \SetCell[r=2]{c,m}\verttext{Modified} \verttext{constant} & \SetCell[r=2]{c,m} \includegraphics[width=0.05\textwidth]{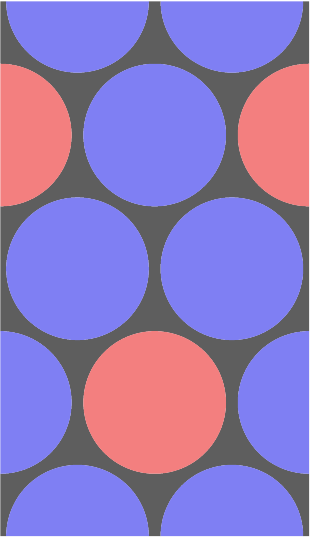} & & & & & &
        \\
        $s_1$ & & & {$\vec{\pi}_3$} & {$-0.97\cdot10^{-3} E$} & {$0.011 E$} & {$0.011 E$} &  \SetCell[c=2]{h} Close to isotropic & & $\hat{f}_\mathrm{bg} \in \langle 1.69, 1.79 \rangle$ & Many small
        \\ \hline
    \end{tblr}
    \caption{Summary of pneumatic actuation schemes, their resulting patterns, critical average pressure differences causing patterning, and resulting values of post-bifurcation macroscopic properties. The normal components of the effective macroscopic stiffness tensor $\bar{D}_{1111}^\mathrm{postbif}$ and $\bar{D}_{2222}^\mathrm{postbif}$ are computed immediately after bifurcation (i.e., patterning). All reported pressures and stiffnesses are normalized by $E$, the reference Young's modulus of the bulk material. The frequency of the acoustic bandgap $\hat{f}$ is normalized according to Equation~(\ref{eq:freqnorm}). The main difference between the schemes is the isotropy of post-bifurcation response and more specifically its direction, expressed here as the angles $\theta_{\bar{D}_{\perp,\mathrm{max}}}$ and $\theta_{\bar{D}_{\perp,\mathrm{min}}}$. Those are measured from the $\vec{e}_1$ axis and represent the angle at which the normal component of macroscopic effective stiffness is maximal and minima, respectively. Furthermore, the acoustic properties differ. The modified constant scheme $s_1$, which triggers the symmetry-conserving deformation pattern $\vec{\pi}_3$, maintains an acoustic resonance bandgap during pneumatic loading, unlike the other schemes. Upon bifurcations, different acoustic bandgaps open for each scheme.}
    \label{tab:rvesummary}
\end{sidewaystable}


\section{Experimental validation}
\label{sec:experiments}

Given the sensitivity of bifurcation to imperfections, we experimentally validated that the numerically obtained results are physically viable. This verification has been performed on finite samples of two different sizes, with either 19 or 61 voids, see \hyperref[sec:experimentalmethods]{Methods} for a detailed discussion.

\subsection{Small sample}

Numerical analyses showed that the small sample with 19 voids could be actuated in the row-wise scheme $s_2$ to form pattern $\vec{\pi}_2$, or in a constant scheme $s_0$ to form pattern $\vec{\pi}_3$.

\begin{figure}[p]
    \centering
    \includegraphics{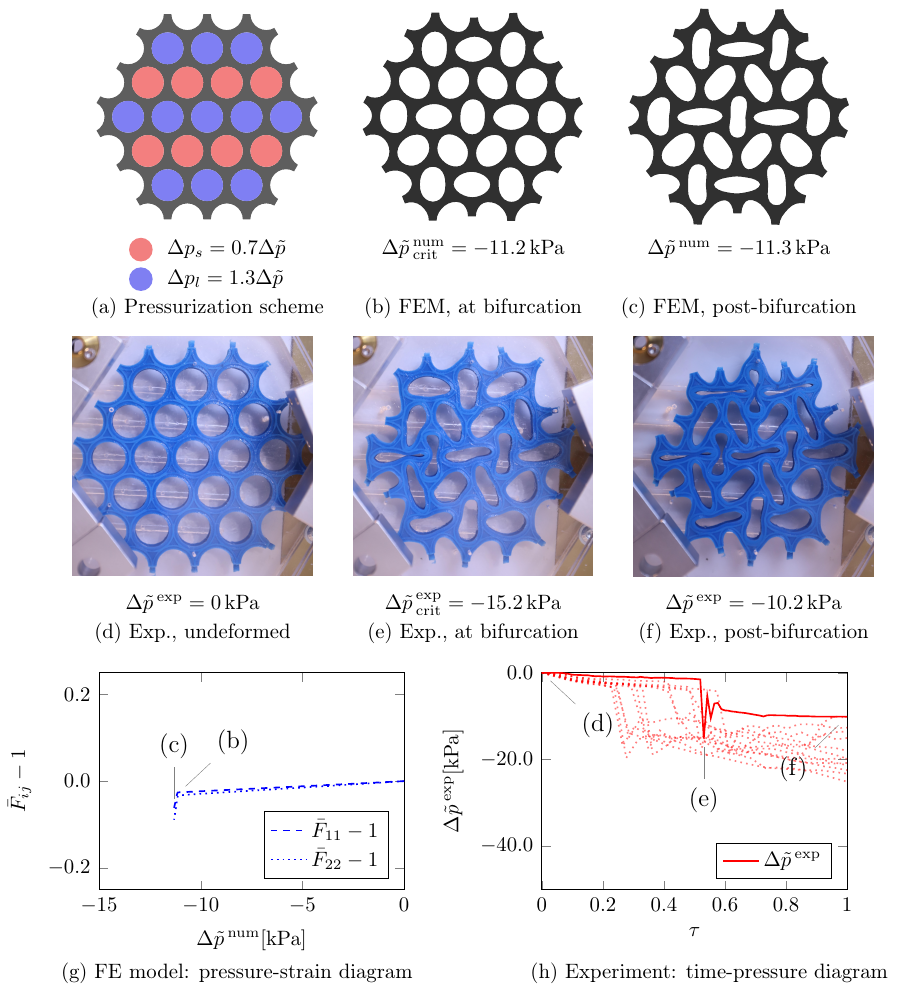}
    \caption{Simulation and experimental results of the \textbf{row-wise pressurization scheme $s_2$} applied to a \textbf{small} finite sample, leading to the \textbf{butterfly pattern} $\vec{\pi}_2$. (a)~Undeformed configuration of  the finite element model and the pressurization scheme. (b-c)~Deformed configurations of the finite element model for different levels of the average pressure difference $\Delta \tilde{p}\,^\mathrm{num}$. (d-f)~Deformed states of the sample in experiment at different load levels $\Delta \tilde{p}\,^\mathrm{exp}$. (g)~Evolution of selected components of the average strain tensor $\bar{\bm{F}}$ with the pneumatic load in the finite element simulations, where the sharp corners mark the critical buckling load $\Delta \tilde{p}\,^\mathrm{num}_\mathrm{crit}$. (h)~Evolution of the average pressure difference $\Delta \tilde{p}\,^\mathrm{exp}$ as measured on the air pumps, with relative experiment time $\tau \in \langle 0, 1 \rangle$. At the bifurcation point, the experimental apparatus begins to leak air with the rapid deformation of the sample, resulting in temporary pressure loss which reveals in the graph the critical load $\Delta \tilde{p}\,^\mathrm{exp}_\mathrm{crit}$. Plots of multiple experiments are shown; the one documented by the photos above is plotted in the thick solid line.}
    \label{fig:smallresult_rows}
\end{figure}

For the row-wise $s_2$ actuation of the small sample, both simulation and experiment lead to patterning into the butterfly pattern $\vec{\pi}_2$ (the orientation of the pattern differs), shown in Figure~\ref{fig:smallresult_rows}bcef. However, the simulation identified a critical load of  $\Delta \tilde{p}\,^\mathrm{num}_\mathrm{crit} = \SI{-11.2}{\kilo\pascal}$, while the experiments revealed a decidedly higher critical loads on average, in the range of $\Delta \tilde{p}\,^\mathrm{exp}_\mathrm{crit} = \langle -12.1, -19.7 \rangle \; \si{\kilo\pascal}$. The response of the experimental sample thus appears to be significantly stiffer than predicted numerically, which can be attributed to differences between the numerical and experimental setups. Several uncertainties and phenomena had not been included in the numerical model, which is a plane strain finite element simulation of a simply supported finite sample. In the experiment, conversely, air leakage is prevented by the sample being weighed down by the top plexiglass sheet and even additional weights, thus introducing out-of-plane loading. Care has been taken to reduce this additional weight to the necessary minimum, yet still certain effects persist. Firstly, out-of-plane compression is introduced to the sample. Additionally, friction between the sample and the plexiglass also plays a role, despite extensive lubrication. We have tested this assumption by performing several experiments (not included among those in Figure~\ref{fig:smallresult_rows}h for consistency) with a larger weight placed on the plexiglass. With increasing out-of-plane load, a trend towards a higher load level of the critical pressure difference has been confirmed.

\begin{figure}[p]
    \centering
    \includegraphics{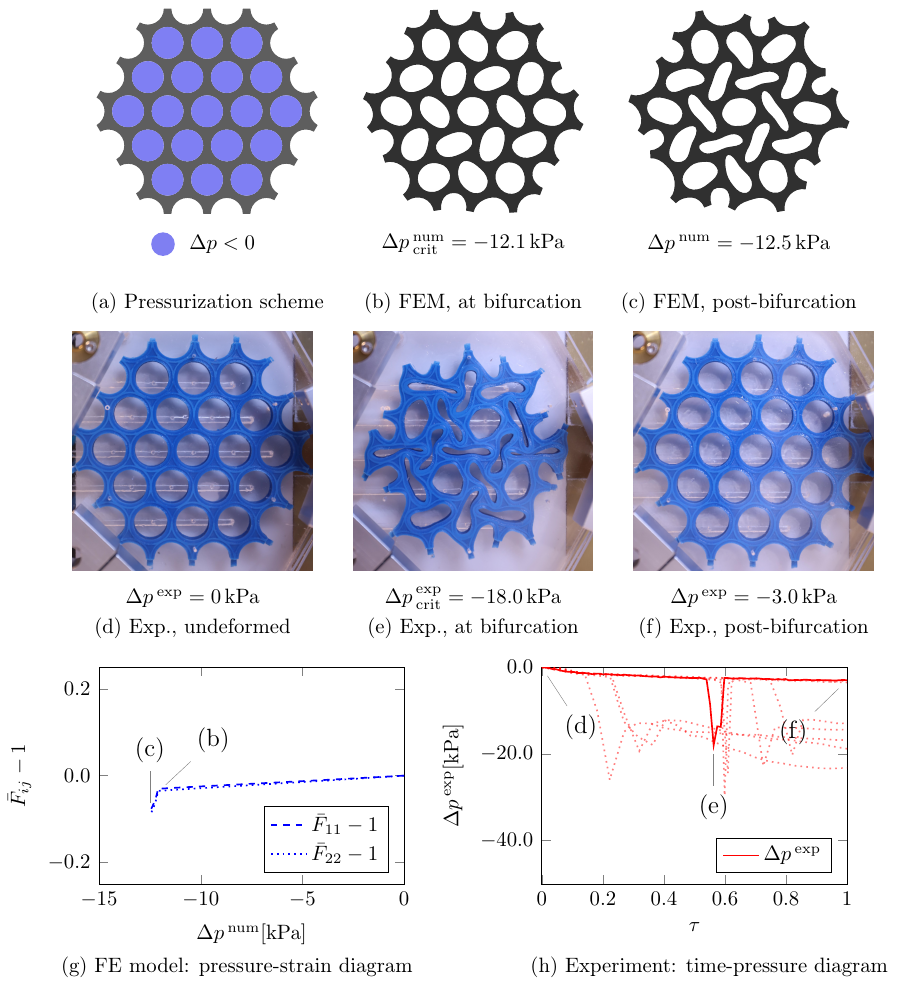}
    \caption{Simulation and experimental results of the \textbf{constant pressurization scheme $s_0$} applied to a \textbf{small} finite sample, leading to the \textbf{flower-like pattern $\vec{\pi}_3$}. (a)~Undeformed configuration of  the finite element model and the pressurization scheme. (b-c)~Deformed configurations of the finite element model for different levels of the introduced pressure difference $\Delta p\,^\mathrm{num}$. (d-f)~Deformed states of the sample in experiment at different load levels $\Delta p\,^\mathrm{exp}$. Note that in the experiment pictured, post-bifurcation leaks due to shifting of the sample resulted in unloading into almost the reference state. (g)~Evolution of selected components of the average strain tensor $\bar{\bm{F}}$ with the pneumatic load in the finite element simulations, where the sharp corners mark the critical buckling load $\Delta p\,^\mathrm{num}_\mathrm{crit}$. (h)~Evolution of actuation pressure difference $\Delta {p}\,^\mathrm{exp}$ measured on air pumps with relative experiment time $\tau \in \langle 0, 1 \rangle$. At the bifurcation point, the experimental apparatus begins to leak air with the rapid deformation of the sample, resulting in temporary pressure loss which reveals in the graph the critical load $\Delta p\,^\mathrm{exp}_\mathrm{crit}$. Plots of multiple experiments shown; the one documented by the photos below is represented by the thick solid line.}
    \label{fig:smallresult_const}
\end{figure}

Unlike on the infinite microstructure, the modification of the constant scheme $s_0$ into the modified scheme $s_1$ to preselect certain voids of the microstructure as the centers of the flower-like pattern is not necessary on the finite structure, as the boundary effects present here remove the uncertainty and the pattern $\vec{\pi}_3$ is formed on the small sample both in simulation (around the central void) and in experiment (around the voids in the inner circle around the center), see Figure~\ref{fig:smallresult_const}. In this case, the simulation predicted a critical pressure difference value of $\Delta p\,^\mathrm{num}_\mathrm{crit} = \SI{-12.1}{\kilo\pascal}$, while the experimental results varied in the range of $\Delta p\,^\mathrm{exp}_\mathrm{crit} = \langle -17.9,-29.6 \rangle \;\si{\kilo\pascal}$. We attribute the difference to the same factors as in the previous case.

\subsection{Large sample}

On the larger sample with 61 voids, simulations predicted that pressurization schemes $s_2$ and $s_3$ are possible, leading to patterns $\vec{\pi}_2$ and $\vec{\pi}_1$, respectively. The flower-like pattern $\vec{\pi}_3$ is not attainable because this sample is no longer small enough for its formation in response to the constant scheme $s_0$ and applying the modified scheme $s_1$ is not supported by the experimental apparatus, see \hyperref[app:apparatus]{Methods}.

\begin{figure}[p]
    \centering
    \includegraphics{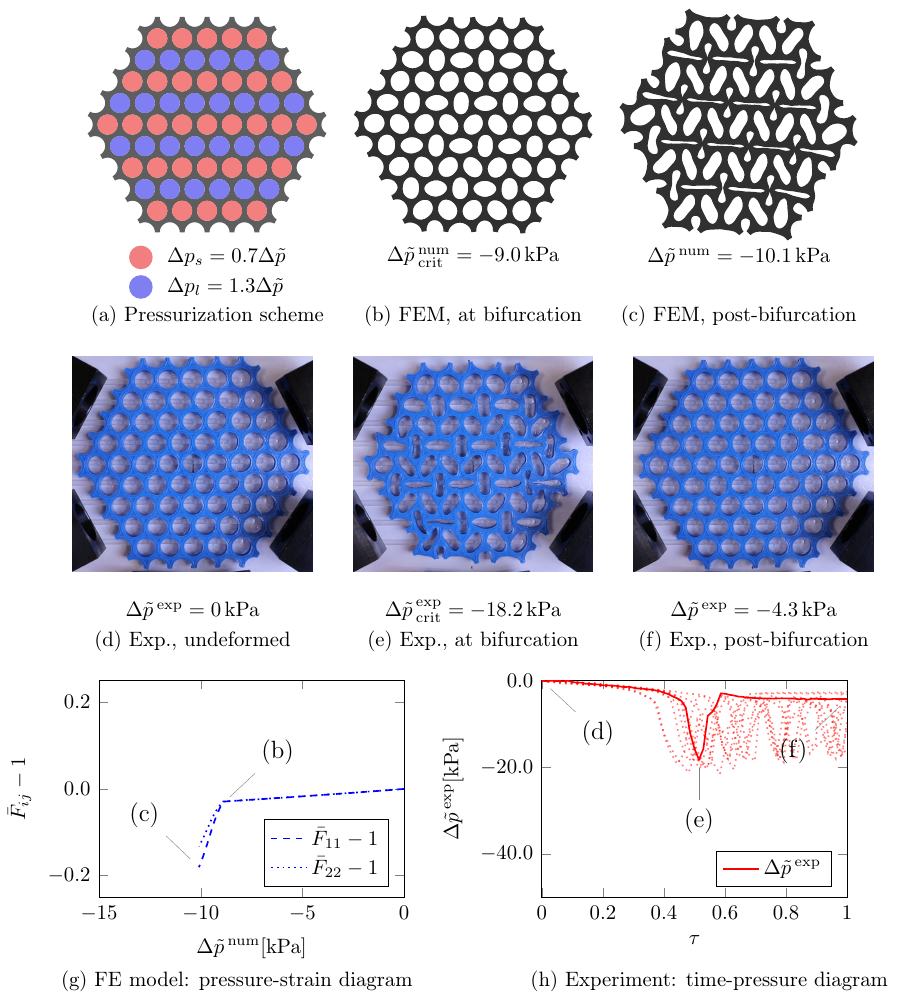}
    \caption{Simulation and experimental results of the \textbf{row-wise pressurization scheme $s_2$} applied to a \textbf{large} finite sample, leading to the \textbf{butterfly pattern $\vec{\pi}_2$}. (a)~Undeformed configuration of  the finite element model and the pressurization scheme. (b-c)~Deformed configurations of the finite element model for different levels of the average pressure difference $\Delta \tilde{p}\,^\mathrm{num}$. (d-f)~Deformed states of the sample in experiment at different load levels $\Delta \tilde{p}\,^\mathrm{exp}$. Note again the unloading due to pressure leaks. Parts of weights holding down the plexiglass are visible in the photos. (g)~Evolution of selected components of the average strain tensor $\bar{\bm{F}}$ with the pneumatic load in the finite element simulations, where the sharp corners mark the critical buckling load $\Delta \tilde{p}\,^\mathrm{num}_\mathrm{crit}$. (h)~Evolution of the average pressure difference $\Delta \tilde{p}\,^\mathrm{exp}$ as measured on the air pumps, with relative experiment time $\tau \in \langle 0, 1 \rangle$. At the bifurcation point, the experimental apparatus begins to leak air with the rapid deformation of the sample, resulting in temporary pressure loss which reveals in the graph the critical load $\Delta \tilde{p}\,^\mathrm{exp}_\mathrm{crit}$. Plots of multiple experiments are shown; the one documented by the photos above is plotted in a thick solid line. In some of the experiments, repeated loading and unloading was caused by pressure leaking out in the deformed state and those leaks closing again with unloading.}
    \label{fig:largeresult_rows}
\end{figure}

The row-wise scheme $s_2$ as applied to the large sample is depicted in Figure~\ref{fig:largeresult_rows}. Similarly to the smaller sample, the expected butterfly pattern $\vec{\pi}_2$ appears both in the simulation as well as in the experiment. In both cases the pattern is more pronounced in the center of the sample. This is due to boundary effects and, in the experiment, possibly even due to sample-plexiglass friction. The simulation predicted a critical pressure difference of $\Delta \tilde{p}\,^\mathrm{num}_\mathrm{crit} = \SI{-9.0}{\kilo\pascal}$ with the experimental results in the range of $\Delta \tilde{p}\,^\mathrm{exp}_\mathrm{crit} = \langle -18.2,-21.6 \rangle \;\si{\kilo\pascal}$. This is an even more pronounced discrepancy than that of the small sample. Presumably this is caused by two factors: firstly, the weight placed on the plexiglass to prevent pressure leaks had to be increased with the load per area increasing approximately $2.5$ times; secondly, the larger sample size leads to larger absolute values of displacements due to patterning, resulting in a necessity to overcome more friction resistance of the weighted plexiglass.

\begin{figure}[p]
    \centering
    \includegraphics{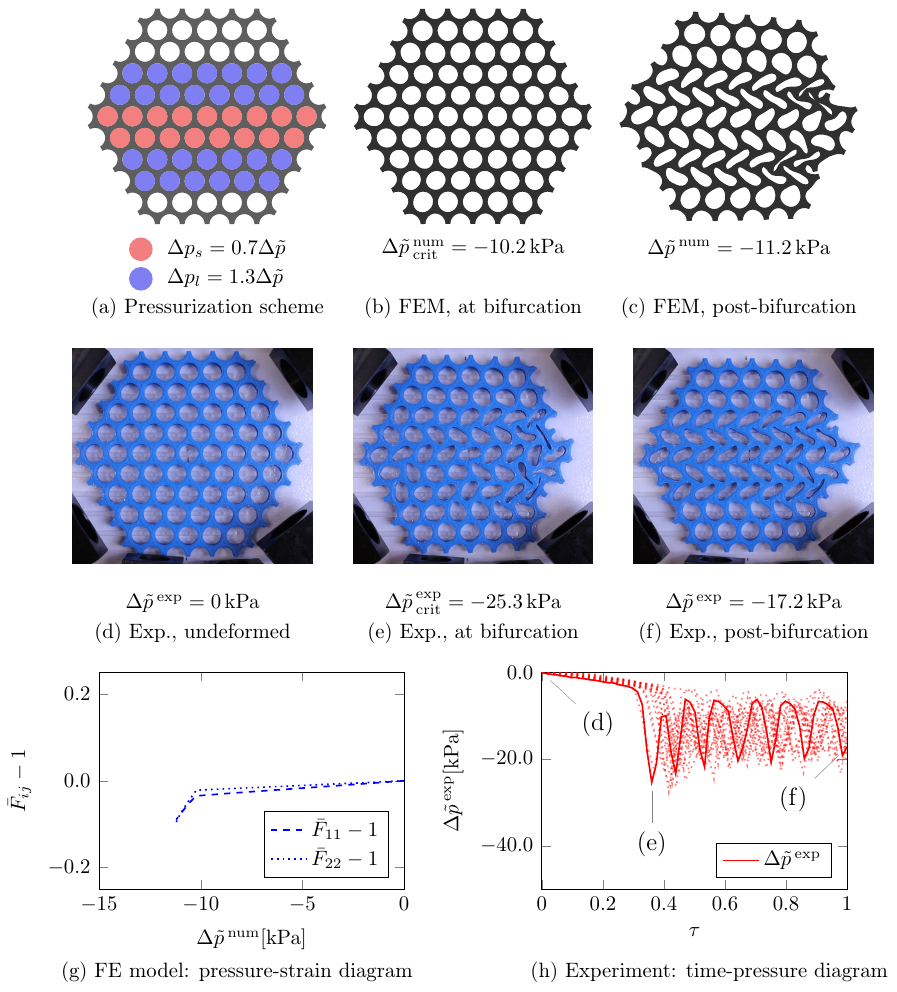}
    \caption{Simulation and experimental results of the \textbf{double-row-wise pressurization scheme $s_3$} applied to a \textbf{large} finite sample. Boundary rows of voids are left unpressurized to avoid undesired boundary effects. This scheme leads to the \textbf{shear pattern $\vec{\pi}_1$}. (a)~Undeformed configuration of  the finite element model and the pressurization scheme. (b-c)~Deformed configurations of the finite element model for different levels of the average pressure difference $\Delta \tilde{p}\,^\mathrm{num}$. (d-f)~Deformed states of the sample in experiment at different load levels $\Delta \tilde{p}\,^\mathrm{exp}$. Similarly to the row-wise scheme, repeated loading and unloading is observed due to pressure leaks. The pattern improves upon subsequent triggering. Parts of weights holding down the plexiglass are visible in the photos. (g)~Evolution of selected components of the average strain tensor $\bar{\bm{F}}$ with the pneumatic load in the finite element simulations, where the sharp corners mark the critical buckling load $\Delta \tilde{p}\,^\mathrm{num}_\mathrm{crit}$. (h)~Evolution of the average pressure difference $\Delta \tilde{p}\,^\mathrm{exp}$ as measured on the air pumps, with relative experiment time $\tau \in \langle 0, 1 \rangle$. At the bifurcation point, the experimental apparatus begins to leak air with the rapid deformation of the sample, resulting in temporary pressure loss which reveals in the graph the critical load $\Delta \tilde{p}\,^\mathrm{exp}_\mathrm{crit}$. Plots of multiple experiments are shown; the one documented by the photos above is plotted in a thick solid line.}
    \label{fig:largeresult_doublerows}
\end{figure}

In the case of the double-row-wise pressurization scheme $s_3$, a modification was necessary on the large finite sample which left some of the border rows without pressurization, see the scheme pictured in Figure~\ref{fig:largeresult_doublerows}a. This is due to observed boundary effects: with an odd number of rows in the sample, pressurizing the boundary rows leads to different patterns being triggered locally and thus interfering with the deformation of the sample into the desired shear pattern $\vec{\pi}_1$. When the boundary rows are left unpressurized, however, the pattern eventually propagates from the center of the sample, where it is triggered as expected. The results of the $s_3$ scheme are pictured in Figure~\ref{fig:largeresult_doublerows}. The finite element simulation predicted a critical pressure difference of $\Delta \tilde{p}\,^\mathrm{num}_\mathrm{crit} = \SI{-10.2}{\kilo\pascal}$, while the experiments lay in the range of $\Delta \tilde{p}\,^\mathrm{exp}_\mathrm{crit} = \langle -19.2,-27.6 \rangle \;\si{\kilo\pascal}$. The unpressurized voids have not been included in the determination of the average pressure differences here.

Admittedly, the deformed shapes of the experimental samples in all the presented results are not as neat as the results of the finite element simulations due to bespoke experimental uncertainties; there is certain noise on top of the recognizable patterns, which we attribute to numerous imperfections of the material and of the loading setup, chief among which might be the uneven lateral load causing non-uniform friction effects across the sample. Nevertheless, the known bifurcation patterns are still visibly discernible and can be repeatedly and reliably triggered by the designed actuation schemes. This leads us to the conclusion that the experiments validate the ability of these schemes to trigger patterns within the microstructure. We wish to highlight that for each of the two sample sizes all experiments have been performed on a single rubber sample. With care taken to unload to a truly stress-free configuration, repeated loading and unloading in different schemes has not been observed to have any measurable effect on the samples' behavior, corroborating the argument of the patterning process being reliably reversible. Although the material itself is surely viscoelastic under large strains and although friction introduces plasticity into the system, patterns can be reproducibly triggered.

\section{Conclusions}
\label{sec:conclusions}

We presented non-uniform pressure actuation as an alternative to external macroscopic strain loading for triggering internal instabilities in pattern forming mechanical metamaterials. More specifically, we focused on 2D honeycomb microstructures with hexagonal stacking of circular voids, which are known to exhibit three distinct internal patterns upon macroscopic compression: the shear, the butterfly and the flower-like pattern. Our goal had been to trigger the same patterns pneumatically. The proposed non-uniform actuation schemes proved to be capable of steering the microstructure towards all three patterns, including the three modes of the shear pattern $\vec{\pi}_1$ and similarly three different orientations of the butterfly pattern $\vec{\pi}_2$, i.e., seven distinct configurations in total. We have confirmed this by numerical simulations of periodically repeating infinite microstructures and, for several actuation schemes transferred to finite samples, experimentally as well. For the experiment, we cast the finite size samples from silicone rubber in a 3D printed mold and subjected them to pneumatic actuation in an in-house built apparatus capable of introducing different pressures to different rows of voids in the samples. The samples had been encased inbetween lubricated plexiglass sheets to mimic assumed plane strain conditions and to prevent air leakage. While this setup caused stiffer behavior than predicted numerically, it nevertheless allowed to qualitatively prove that the chosen schemes indeed lead to the three patterns in reality as well as in simulation.

Furthermore, with the simulations of periodic microstructures we showcased the impact of the patterns formed by the proposed actuation on macroscopic stiffness. Not only does the triggered patterning process lead to a reduction of macroscopic stiffness, it also influences the anisotropy of the macroscopic behavior. Thus the present actuation method enables a single microstructure to change the directionality of its response to macroscopic loading in an independent manner. In this way, this research paves the way towards active mechanical metamaterial designs exploiting such behavior in the context of soft robotics, shape-morphing metamaterials, or pneumatically actuatable dampers.

In addition to the stiffness tuning, we also demonstrated the use of pneumatically actuated honeycomb microstructures for active tuning of acoustic material properties. The three proposed actuation schemes lead to different acoustic responses during the loading cycle. A resonance bandgap present in the reference configuration of the microstructure is preserved in the pre-bifurcation regime only when the $s_1$ scheme is applied and closes in other cases. Different acoustic bandgaps open in the post-bifurcation regime for the different actuating schemes and deformation patterns. Thus the pneumatic actuation in the different schemes can be utilized to actively select the resonance bandgaps desired at the moment. Active acoustic wave insulators or manipulators based on patterning metamaterials are known to be viable concepts in literature~\cite{Liu2022squarelatticecoatingbandgaps, Bertoldi2008bandgapsinpatterning, Shan2014elasticwavesinhexagons}.

This work therefore serves as a proof-of-concept raising several open questions. Most notably, we have focused only on hexagonal lattices with uniformly sized circular voids and on triggering the known modes and patterns. More patterns may be identified with further, possibly more complicated pneumatic actuation schemes. For microstructures without pneumatic actuation, this has been achieved by analysis of symmetry groups of both loading and microstructural geometry~\cite{Azulay2024predictingpatterns, Hendriks2024GNNsymmetries, hendriks2025WallpaperGroupDatasetPreprint}. Another apparent next step would be an analysis of non-uniform pneumatic actuation on microstructures with more involved design, leading to other interesting behavior such as bistable or multistable structures, where pneumatic pressure is used to switch between phases. Last but not least, more development towards applications of the concepts introduced herein is necessary, possibly moving towards 3D structures.


{\footnotesize
\section*{Methods}

\subsection*{Numerical simulations}
\label{sec:numerical_methodology}

We use non-uniform pressure actuation schemes to trigger different patterns in the same initial microstructure. Since the pattern transformation of honeycombs is influenced by boundary effects in finite samples \cite{Ameen2018sizeeffectonhoneycombs}, we study at first a representative volume element (RVE) constrained by periodic boundary conditions. The results, therefore, are indicative of behavior of an infinitely large slab of material. We focus on two-dimensional structures here, with the assumptions of plane strain behavior and of a hyperelastic material response.

Two different two-dimensional RVEs are used to capture the periodicity of the patterns studied. Their geometry and boundary conditions are pictured in Figure~\ref{fig:rve_geometry}. Since the geometry of the hexagonal lattice is only described by two parameters (the hexagon outer diameter $L$ and the void diameter $d$), the choice of the RVE size is mainly dictated by the periodicity of the applied pressurization schemes. A rectangular RVE consisting of 8 voids (shown in Figure \ref{fig:rve_geometry}a) is sufficient for most of the studied schemes, although a hexagonal RVE could conveniently be used as well. For schemes with incompatible actuation wavelength, a twice larger RVE of 32 voids (pictured in Figure \ref{fig:rve_geometry}b) has to be chosen.

The simulations are performed using an in-house MATLAB~\cite{MATLAB2023} and C++ codebase based on the geometrically nonlinear finite element method. For meshing, the GMSH open-source software is used~\cite{Geuzaine2009gmsh}. Two-dimensional plane strain behavior is assumed; it corresponds to a behavior of a layer in thicker block of material, and partially also to the experimental conditions described in Section~\ref{sec:experiments}. The RVEs are meshed using quadratic triangular finite elements with maximum symmetry of the mesh. Approximating the response of silicone rubber, the neo-Hookean hyperelastic material law~\cite{rivlin1948neohookean} is used to describe the bulk material, with its strain energy density defined as

\begin{equation}
    \label{eq:neohookeanenergydensity}
    W_{\mathrm{NH}} = \frac{\mu}{2} \left(I_1  - 3 -2\log{J}\right) + \frac{\lambda}{2}\left( J - 1\right)^{2}
\end{equation}

\noindent where $I_1 = \mathrm{tr}(\bm{F}^{\mathrm{T}}\bm{F})$ denotes the first invariant of the right Cauchy-Green deformation tensor $\bm{C} =\bm{F}^\mathrm{T}\bm{F}$ with $\bm{F}$ the deformation gradient, and $J = \det{\bm{F}}$ is the deformation gradient Jacobian. The material parameters $\lambda$ and $\mu$ correspond in the small strain limit to the standard Lamé's coefficients. We assume near incompressibility for silicone rubber with the Poisson's ratio $\nu=0.499$ \cite{Mott2009Poissonrationvalues}. This leaves only the Young's modulus $E$ as a free material parameter, since

\begin{equation}
\label{eq:lamecoeffs}
    \lambda = \frac{E\nu}{(1-2\nu)(1+\nu)} \quad\quad \mu = \frac{E}{2(1+\nu)} 
\end{equation}

\noindent For the RVE simulations, we present all results in a dimensionless format, with all stress-like and stiffness-like quantities normalized by the Young's modulus $E$.

\begin{figure}[t]
    \centering
    \begin{tabular}{c c}
         \includegraphics{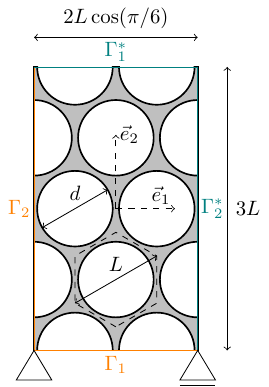} & \includegraphics{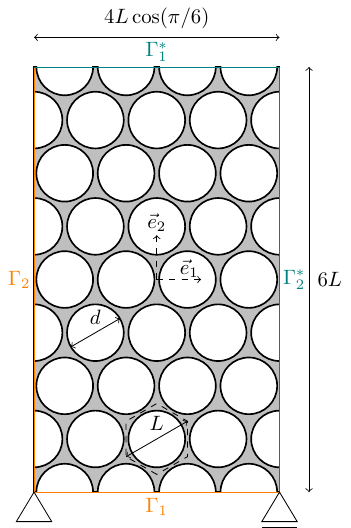}
         \\
         (a) & (b) 
    \end{tabular}
    
    \caption{The geometry of representative volume elements used for numerical simulations of infinite microstructures. (a) The basic RVE consisting of 8 voids. (b) A doubled RVE consisting of 32 voids, necessary for pressurization schemes which do not fit into the basic shape. The geometry of the hexagonal lattice in the $(\vec{e}_1, \vec{e}_2)$ coordinate system is determined entirely by the hexagon outer diameter $L$. The diameter of the voids is chosen as $d=0.8L$. The RVE is fixed against rigid body motions by Dirichlet boundary conditions in the lower corners as pictured; it is, however, not fixed against global buckling modes of the microstructure. Periodic boundary conditions are imposed to make the boundaries $\Gamma_1^*$ and $\Gamma_2^*$ (teal) periodic images of the boundaries $\Gamma_1$ and $\Gamma_2$ (orange), respectively.}
    \label{fig:rve_geometry}
\end{figure}

To simulate the pneumatic actuation, we employed a third medium method developed by Faltus~et~al.~\cite{Faltus20243M}. It relies on a fictitious material inside the voids to introduce externally prescribed hydrostatic Cauchy stress, which also stiffens upon extreme compression to enforce frictionless contact constraints. The model also introduces regularization terms to prevent excessive distortions of the fictitious mesh; those are, however, typically not necessary for the simulations included in this study. The third medium material is thus characterized by the strain energy density of

\begin{equation}
    W_\mathrm{3M}  = \gamma K_\mathrm{ch}\ln^2{J} + \gamma K_\mathrm{ch}\left(J^{-2/3}I_1  - 3\right) + \Delta p J
\end{equation}

\noindent where $\gamma \approx 10^{-6}$ is the third medium stiffness reduction coefficient, $K_\mathrm{ch}$ is a characteristic bulk material stiffness, and $\Delta p$ is the prescribed pneumatic pressure, expressed as the difference with respect to the atmospheric pressure.

To handle instabilities and bifurcations that inherently appear in simulations of pattern-forming metamaterials, we used a modified Newton solver based on the Cholesky decomposition algorithm, which directs the solver away from local maxima and saddle points of the energy landscape \cite{Faltus20243M, Cheng1998modifiedcholesky, Nocedal2006numericaloptimization}. This allows finding stable bifurcation branches, albeit at the cost of an increased number of solver iterations in the affected solution step.

Effective macroscopic properties such as macroscopic strain, macroscopic stress and stiffness have been obtained from the microstructural results through first order numerical homogenization~\cite{Kouznetsova2001approach}.

For the strain loading simulations presented in Sections~\ref{sec:rveproperties} and \ref{sec:bloch}, the same RVE from Figure~\ref{fig:rve_geometry}a was used, albeit without the third medium mesh. The macroscopic strain load was imposed as a prescribed displacement of the RVE corners. For the calculation of the stiffness anisotropy curve of pattern $\vec{\pi}_3$ in Figure~\ref{fig:isotropy}b, a special hexagonal RVE with full rotational and mirror mesh symmetry was used, adapted from Rokoš~et~al.~\cite{Rokos2020honeycombs}. This way, the precise isotropy of the response could be captured as exactly as possible.

\subsection*{Dispersion analysis}
\label{app:bloch}

The dispersion analysis presented in Section~\ref{sec:bloch} was performed numerically in an in-house MATLAB code using methodology largely adopted from Cool~et~al.~\cite{cool2024dispersionfordummies}, Ning~et~al.~\cite{ning2021bandgapsinmetamatswithinclusions} and Maurin~et~al.~\cite{maurin2018bandgapextremumonibz}.

For each deformed state of each periodic RVE from finite element simulations (see \nameref{sec:numerical_methodology} and Figure~\ref{fig:rve_geometry}), the Bloch boundary conditions were applied to study the free propagation of different waves. First, the unconstrained tangential stiffness matrix $\mathsf{K}$ and mass matrix $\mathsf{M}$ were constructed on the deformed state of the RVE. For the construction of $\mathsf{M}$, a uniform mass density of $\rho = 1$ was applied, since the bulk material of the unit cell is considered homogeneous and the normalized frequency result is independent of the mass density (see ahead Equation~(\ref{eq:freqnorm})). In the case of pneumatic actuation, the third medium material~\cite{Faltus20243M} was present in the voids. To prevent influence of nonphysical oscillations within this fictitious material on the solution, the degrees of freedom associated with the void mesh were removed using static condensation, thus eliminating them from the analysis yet preserving the effect of pneumatic loading within $\mathsf{K}$. Since the third medium is a fictitious material, the mass associated with the condensed degrees of freedom was ignored.

A dispersion diagram relates wave vectors $\vec{k}$ to their frequencies. Such diagrams were obtained by numerically solving the generalized dispersion eigenvalue problem~\cite{cool2024dispersionfordummies}

\begin{equation}
\label{eq:gdep}
    (\tilde{\mathsf{K}} - \omega^2\tilde{\mathsf{M}})\tilde{\mathsf{q}} = \mathsf{0}
\end{equation}

\noindent for the eigenvalues $\omega^2$ and eigenvectors $\tilde{\mathsf{q}}$. Here $\tilde{\mathsf{K}}$ and $\tilde{\mathsf{M}}$ were obtained from $\mathsf{K}$ and $\mathsf{M}$ as

\begin{equation}
    \tilde{\mathsf{K}} = \mathsf{R}^T\mathsf{K}\mathsf{R} \quad\quad \tilde{\mathsf{M}} = \mathsf{R}^T\mathsf{M}\mathsf{R}
\end{equation}

\noindent where $\mathsf{R}(\vec{k})$ is the periodicity matrix, which depends on a single chosen wave vector $\vec{k}$ defining the wave, see Cool~et~al.~\cite{cool2024dispersionfordummies} for further details.

The results of the generalized eigenvalue problem~(\ref{eq:gdep}) were converted to normalized eigenfrequencies $\hat{f}$ according to

\begin{equation}
\label{eq:freqnorm}
        \hat{f} = \frac{\omega L_\mathrm{char}}{2\pi c_\mathrm{char}}
\end{equation}

\noindent where $\omega$ is the angular eigenfrequency, a result of solving~(\ref{eq:gdep}), $L_\mathrm{char}$ is the characteristic size of the RVE, taken as one half of the deformed RVE height in our case, and $c_\mathrm{char} = \sqrt{\mu/\rho}$ is the characteristic wave speed equal to the square root of the ratio of the bulk material shear modulus $\mu$ from Equation~(\ref{eq:neohookeanenergydensity}), and the bulk material mass density $\rho$. The dispersion analysis has been performed to cover the eigenfrequency range of $\hat{f} \in \langle 0, 4 \rangle$. Since we considered only real vectors for $\vec{k}$ (i.e., freely propagating waves) and our unit cell was undamped, all resulting eigenfrequencies are also real. 

The considered wave vectors $\vec{k}$ were chosen on the boundary of the Irreducible Brillouin Zone (IBZ), which was identified based on the symmetries of the unit cell according to the procedure presented in Maurin~et~al.~\cite{maurin2018bandgapextremumonibz}. The IBZ for the reference configuration of the unit cell and for the deformed configurations under macroscopic strain loading is rectangular. In case of the pneumatic loading, the RVE was not constrained against global shear deformation, the presence whereof would require an oblique IBZ. Since, however, only a negligible macroscopic shear appeared in the actual deformed configurations, we used a rectangular IBZ for these cases as well. To verify the validity of the assumption that the the extrema of the dispersion surfaces lie on the IBZ contour, which is not definitively proven~\cite{cool2024dispersionfordummies,maurin2018bandgapextremumonibz}, we constructed the full dispersion surface in a few selected configurations, finding no difference beyond the magnitude of a numerical error in the identified extrema. Hence, the dispersion diagrams are constructed by sweeping the IBZ contour only.

Solving the generalized eigenvalue problem~(\ref{eq:gdep}) for each $\vec{k}$ at the IBZ contour thus reports eigenfrequencies $\hat{f}$ at which waves freely propagate through the microstructure. Frequency ranges at which waves propagate for no $\vec{k}$ represent acoustic bandgaps; they are reported as the blue bands in Figure~\ref{fig:referencedispersion} and in Figure~\ref{fig:blochresults}, right column. Since in the numerical approach the dispersion curves are solved point-wise, fake bandgaps can be misidentified at a point where the dispersion curves cross each other~\cite{maurin2018bandgapextremumonibz}. To avoid this, we constructed the dispersion curves by eigenmode tracking between neighboring sampling points of $\vec{k}$, thus capturing their crossing over each other. On top of that, we imposed a numerical tolerance for a minimal bandgap width.

\subsection*{Experiments}
\label{sec:experimentalmethods}

For experimental validation, the infinite samples considered in Section~\ref{sec:simulations} are not feasible, as they would require a structure large enough for at least its central part to be devoid of boundary effects and reasonably close in behavior to a theoretical infinite limit. The size of such a structure would quickly grow impractical, especially considering the need to introduce pneumatic actuation to every void. For this reason, we opted to find samples of feasible size by performing numerical simulations on their actual geometry. With the exception of the periodic boundary conditions, the numerical methodology remained the same for this purpose as for the infinite microstructures, see \nameref{sec:numerical_methodology}. Two different sizes of samples have been identified: one with 19 voids (referred to as the \textit{small sample}) and the other, larger, with 61 voids (the \textit{large sample}). Both geometries are displayed in Figure~\ref{fig:samples} as photos of the actual samples. The spacing of the hexagonal lattice is $L=\SI{25}{\milli\meter}$, recall Figure~\ref{fig:rve_geometry} for its definition. For practical purposes of manufacturing, a smaller void size, $d=0.7L=\SI{17.5}{\milli\meter}$, has been chosen. Small ridges in the sample surfaces, visible in Figure~\ref{fig:samples}, were introduced to improve adhesion to plexiglass around the voids, helping to prevent air leakage.

\begin{figure}[t]
    \centering
    \includegraphics{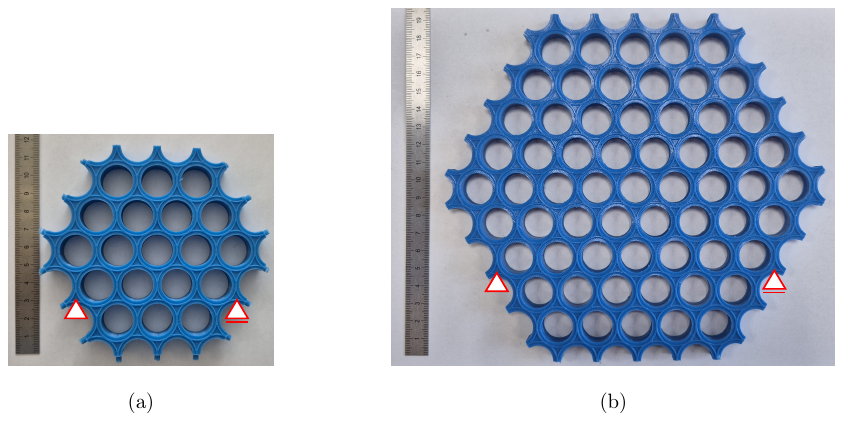}
    \caption{Finite size samples for experimental validation (with an illustration of boundary conditions used in simulations). In both samples, the hexagonal lattice has the size of $L=\SI{25}{\milli\meter}$ with voids of the diameter $d = 0.7L$. The thickness of the rubber samples is $t=\SI{10}{\milli\meter}$. (a) Small 19-void sample. (b) Large 61-void sample.}
    \label{fig:samples}
\end{figure}

The samples were cast from two-phase silicone rubber~\cite{zhermackza22silicone} in a mold 3D printed on a Prusa MK3S printer~\cite{prusa2019mk3smanual} from Prusament PETG filament~\cite{prusa2022prusamentpetgtechsheet}. The mold was designed to minimize formation of air cavities and sample inhomogeneities. From a series of uniaxial compression tests on cylindrical rubber samples conducted on a open-hardware Thymos loading frame~\cite{porubsky2025thymos} and from the assumption of incompressibility ($\nu=0.499$), the Young's modulus for the neo-Hookean material model, recall Equations~(\ref{eq:neohookeanenergydensity}) and (\ref{eq:lamecoeffs}), has been identified as $E = \SI{0.643}{\mega\pascal}$ for the small sample and $E = \SI{0.554}{\mega\pascal}$ for the large sample. The difference is due to the samples being cast from different batches of silicone at different times; the uniaxial tests were conducted on simultaneously cast cylinders for each case.

For the pneumatic actuation we used an in-house built apparatus where the rubber sample was placed inbetween two plexiglass sheets. The plexiglass had been lubricated by Coyote Silkal 93 silicone oil~\cite{automax2022coyotesilkal93datasheet} to reduce friction with the sample; tightness against air leakage was achieved by placing weights on the top plexiglass sheet. Air suction was introduced through holes drilled in the bottom sheet with the possibility to independently tune the level of loading in different rows of voids. The experiment was controlled by tuning the power output of the integrated air pumps. At all times, independent measurements of pressure levels on the two pumps were recorded. Details of the setup are described in \nameref{app:apparatus}. The apparatus managed to keep airtight at least up to the bifurcation point where patterning was triggered, facilitating quantitative determination of the critical load and qualitative determination of the shape of the pattern triggered. Further deformation sometimes led to displacements that caused blocking of pressure channels, or conversely their opening to atmospheric pressure; for this reason the apparatus was not reliably able to maintain the actuation scheme beyond the bifurcation point, sometimes causing oscillating behavior when pneumatic unloading due to air leakage led to renewed airtightness and thus a new loading cycle.

\subsection*{Experimental apparatus}
\label{app:apparatus}

\begin{figure}[t]
    \centering
    \includegraphics{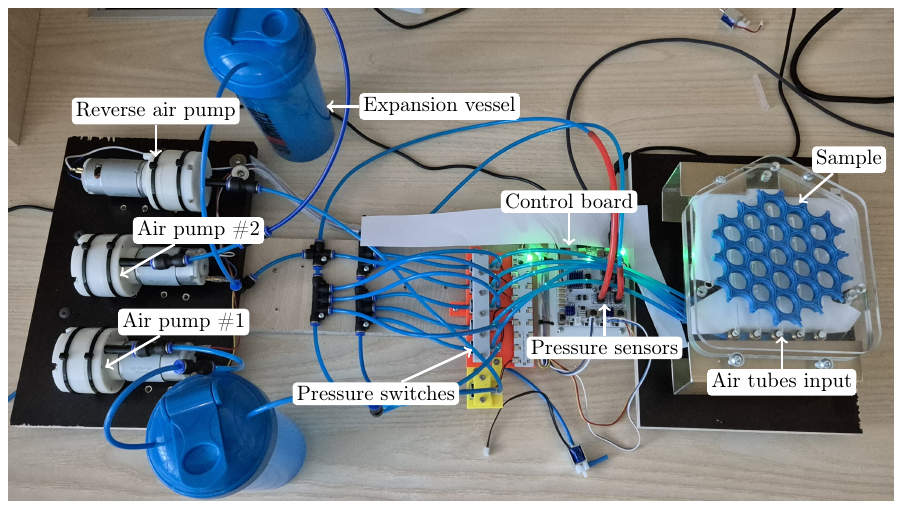}
    \caption{Experimental apparatus for pneumatic actuation of 2D honeycomb structures. The rrubber sample enclosed between plexiglass sheets is mounted next to a control board and connected to 24V air pumps through an array of pressure switches and sensors allowing for regulation and real-time measurement of the pneumatic pressure load in the voids.}
    \label{fig:apparatus}
\end{figure}

The experiments on the rubber sample were performed using an in-house built experimental apparatus pictured in Figure~\ref{fig:apparatus}, purposely designed for this experiment. It consisted of two independently operated $\SI{12}{\watt}$, $\SI{24}{\volt}$ 555 vacuum air pumps capable of introducing $\SI{-60}{\kilo\pascal}$ of suction pressure with expansion vessels for maintaining separate pressure levels, a reverse pump for on-demand reinflating of the sample, pressure switches allowing for selecting which air pumps connect to which of five available air channels, a control board with pressure sensors, and the sample itself enclosed between plexiglass sheets. The top sheet was $\SI{8}{\milli\meter}$ thick while the bottom sheet was made from two $\SI{8}{\milli\meter}$ layers and thus $\SI{16}{\milli\meter}$ thick. With sufficient weights placed on the top plexiglass, this setup was airtight enough in the pre-bifurcation phase, where the deformation of the sample was not extreme. With bifurcation and the resulting large deformation, the alignment of some voids to their corresponding air holes could be disturbed and leakage of air pressure might have occurred.

Albeit the manufactured rubber sample has 19 voids to be pressurized (see Figure~\ref{fig:samples}a), all the experiments conducted used pressurization schemes where the pressure levels were arranged row-wise (compare Figures~\ref{fig:smallresult_rows}a and \ref{fig:smallresult_const}a). For this reason, it was possible to simplify the construction of the apparatus by using only 5 pressure channels, each connected to a single row. This connection was realized through the bottom plexiglass casing, which had been constructed from two sheets glued together. One sheet had been drilled through with the holes connected to the air intakes on the outer side and to horizontal grooves in the adjacent surface of the other sheet on the inner side. These grooves on the other sheet aligned with drill holes, $\SI{4}{\milli\meter}$ in diameter, placed in positions corresponding to the voids of the sample. This way, each of the five air intakes had been connected to a horizontal channel distributing air to each void in one of the rows.

The five pressurizing channels were connected to the two air pumps through an array of 3-way micro solenoid air valves with binary operation through which it was possible to select one of the two pumps for each row independently. The pumps were controlled from a custom-made printed circuit board with an embedded ESP32S3 microcontroller. Pulse-width modulation (PWM) was employed to regulate the operation of the DC motors in the vacuum pumps. Additionally, the control unit features Honeywell SSCDANV015PAAA5 pressure sensors~\cite{honeywell2011sensor} capable of real-time reading of air pressure in each pressure vessel, i.e., informing about pressure levels in each scheme. The experiment was controlled by a MATLAB~\cite{MATLAB2023} script connecting to the control board, sending pump speed instructions, receiving pressure readings in response, and recording exact timestamps of each reading. The sensor readings were adjusted in post-processing so that the maximal air pressure reading at the start of the experiment is considered to be unpressurized state $\Delta p = 0$.

The photos of the experiments presented in Section~\ref{sec:experiments} were taken by a Canon EOS 6D Mark II single-lens reflect camera, mounted on a tripod and suspended above the experimental apparatus. During each experiment, photos were taken in regular intervals and their timestamps were later aligned in post-processing to the timestamps taken by the MATLAB script controlling the apparatus.

\subsubsection*{Adaptation for larger samples}

\begin{figure}[t]
    \centering
    \includegraphics{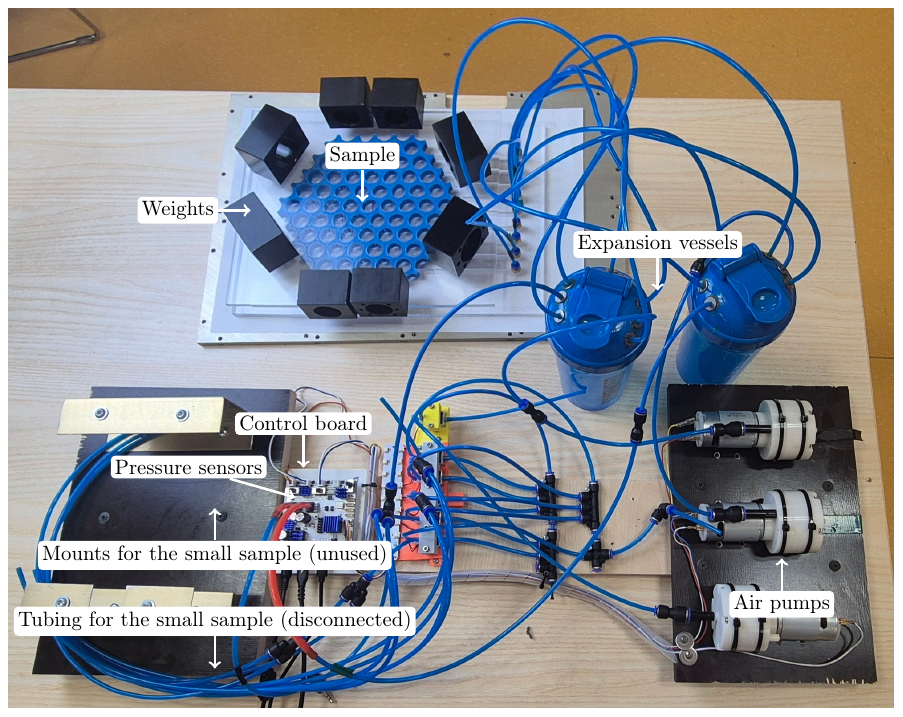}
    \caption{The experimental apparatus adapted for actuating the larger, 61-void sample. The sample is placed outside of the apparatus and the pressure switches are bypassed with the pressurizing scheme being controlled by manual readjustment of the tubing instead.}
    \label{fig:apparatus_large}
\end{figure}

The large 61-void sample required certain adaptations of the experimental setup, see Figure~\ref{fig:apparatus_large}. The construction of the plexiglass sheets has been simplified. The bottom sheet was no longer glued out of two separate sheets, the pressure channels were grooved directly in the lower surface of one sheet and sealed with tape. Maintaining airtightness relied on the stiffness of the underlay the plexiglass sheet was positioned on. This is related to the fact that due to the increased size of the sample, it was no longer mounted directly on the apparatus, but rather positioned on a lab table. The stiff underlay had the benefit of reducing the effect of the imperfect flatness of the plexiglass.

A second significant design change was the bypassing of the pressure switches on the control board. The two pumps were now directly connected to their separate expansion vessels and the switching of pressurization schemes was achieved manually by connecting different expansion vessels to the different inputs on the plexiglass sheet corresponding to each of the nine void rows of the sample (in Figure~\ref{fig:apparatus_large} the configuration for the row-wise scheme $s_2$ can be seen). This modification was necessary to improve the uniformity of the pressure loading across different rows, which was otherwise affected by the differing length of the pressure channels between the pumps and the sample. The pressure sensors on the control board were retained, measuring pressure in the expansion vessels in real time.

}

\clearpage

\paragraph{Acknowledgement}
This work has been supported by the European Union under the ROBOPROX project (reg. no. CZ.02.01.01/00/22 008/0004590). OF acknowledges support of the Czech Science Foundation (grant no.~GA19-26143X) until June 2024, and of the Ministry of Education, Youth and Sport of the Czech Republic (ERC CZ project no. LL2310) from July 2024 onwards. The authors would like to thank Chris Verhoeven, a Master's student at the Eindhoven University of Technology, for his role in finite element code development and numerical simulations, as well as Ondřej Pištora at the Czech Technical University in Prague for his assistance in manufacturing the experimental apparatus.

\paragraph{Declaration of competing interest}
The authors declare that they have no known competing financial interests or personal relationships that could have appeared to influence the work reported in this paper.

\paragraph{Data availability statement}
Data is available in the associated Zenodo repository~\cite{Faltus2025pneumatichoneycombsZenodo} or will be made available upon request.


{\footnotesize
\bibliographystyle{elsarticle-num} 
\bibliography{biblio2.bib}
}




\end{document}